\documentclass[reprint,nofootinbib,preprintnumbers,showkeys,aps,prd]{revtex4-2}
\usepackage{graphicx}
\usepackage{mathtools,amssymb} 
\usepackage{dcolumn} 
\usepackage{bm} 
\usepackage{xcolor}
\usepackage{dsfont} 
\usepackage{slashed} 
\usepackage{hyperref}
\usepackage{mathtools}

\newcommand \beq{\begin{eqnarray}}
\newcommand \eeq{\end{eqnarray}}

\begin{document}
\unitlength=1mm
\allowdisplaybreaks

\title{SU($N_c$) confinement and color $N_c$-ality}

\author{V. Tomas \surname{Mari Surkau}}
\affiliation{Centre de Physique Th\'eorique, CNRS, \'Ecole Polytechnique, IP Paris, F-91128 Palaiseau, France.}

\author{Urko Reinosa}
\affiliation{Centre de Physique Th\'eorique, CNRS, \'Ecole Polytechnique, IP Paris, F-91128 Palaiseau, France.}

\date{\today}

\begin{abstract}
In a recent work, we have argued that the net quark number gained by a bath of quarks and gluons upon bringing an external static quark probe, while being equal to $1$ in the high temperature, deconfined phase, is equal to $0$ or $3$ in the low temperature, confined phase, depending on the value of the quark chemical potential. This establishes a clear-cut connection between the confinement of a medium as probed by order parameters such as the Polyakov loop, and the ability of that same medium to screen the quark probe into states with the same net quark content as mesons or baryons. In this work, we extend these findings to the SU($N_c$) color group, consider color probes in various (fundamental and non-fundamental) representations, and conjecture a rationale of how these results could be recovered from a purely thermodynamic argument that shortcuts the use of the Euclidean functional framework, while emphasizing the role of color $N_c$-ality as a proxy for center symmetry within the Minkowskian framework.
\end{abstract}

\maketitle

\section{Introduction}
Charting the phase diagram of a given system requires the identification of quantities that are prone to behave differently in the various phases of the system, react distinctively to the transition between two such phases or to the nearby presence of particular features such as critical endpoints, triple points, \dots When the difference between the phases is a physically sharp one, this role is embraced by order parameters. In many cases, those actually probe the breaking of a symmetry, thus clearly identifying the physical difference between the two phases. 

In other cases, however, such a difference is more qualitative, and it is thus less obvious to identify proper probes for the different phases. A particular example of this second situation is the physical transition in Quantum Chromodynamics (QCD) as a function of temperature \cite{Aoki:2006we, Borsanyi:2010bp, Aarts2023PhasePhysics}. Numerical simulations of the Euclidean version of this theory, which is relevant for studying the phase diagram, show a continuous change of thermodynamical observables such as the pressure, entropy density, or energy density with the temperature \cite{Borsanyi:2013bia, HotQCD:2014kol}. Still, one refers to this as the confinement/deconfinement transition because, at sufficiently low temperatures, the results can be well accommodated by a hadron gas model, while at high temperatures, the results are more compatible with a gas of quarks and gluons. Thus, although the transition is here a continuous crossover, the system seems to be in two physically distinct phases, which, at least asymptotically, are governed by different degrees of freedom.

To test this hypothesis, one can try to construct quantities that not only probe the transition between the two phases but also give some direct information on the very nature of the active degrees of freedom in each of the phases. Recently, we have proposed a possible candidate for this type of quantity \cite{MariSurkau:2025how}: the net quark number gained by a thermal bath of quarks and gluons upon insertion of an external, static (infinitely heavy) quark probe. Using the simpler setting of heavy-quark QCD (which also features a crossover transition if the quark masses do not exceed a certain value \cite{Fromm2012ThePotentials}), we have shown that this gauge-invariant and RG-invariant quantity,\footnote{Because of these two properties, the net quark number gain qualifies as a good theoretical observable. We are not implying, of course, that it corresponds to an experimental observable since the physical meaning, if any, of bringing a static color probe into the medium still needs to be clarified.} while being equal to $1$ in the high temperature, deconfined phase, is equal to $0$ or $3$ in the low temperature, confined phase, depending on the value of the quark chemical potential. This establishes a clear and direct connection between the confinement/deconfinement transition in heavy-quark QCD and the nature of the degrees of freedom within each of the two phases. The $1$ in the high temperature phase is interpreted as the bath being dominated by quark degrees of freedom, which makes the addition of an external quark increase the net quark number simply by one unit. On the other hand, the $0$ or $3$ in the low temperature phase is interpreted as the bath being dominated by hadronic degrees of freedom, which makes the addition of an external quark only possible if it is absorbed into a meson or a baryon. We have also argued that something similar should occur in QCD when the temperature becomes sufficiently smaller than the constituent quark mass. 

In this work, we extend these findings in various directions, while remaining in the simpler setting of heavy-quark QCD. First, we consider the generalization of the color group to SU$(N_c)$ and show that, in the case where the low-temperature, confining medium is probed with a charge in a fundamental color representation $\nu$ obtained by anti-symmetrization of the color state of $\nu$ quarks, the medium brings out either $\nu$ antiquarks or $\smash{N_c-\nu}$ quarks, which we interpret as the medium screening the color probe into a state with the same net quark content than $\nu$ mesons or one baryon, respectively. Our calculation does not allow us to access the color representation of the formed states, but it clearly shows that those states have a vanishing $N_c$-ality, a property shared by color-singlet states. What state is created depends on the quark chemical potential $\mu$. Writing 
\beq
\mu_\nu\equiv \left(1-\frac{2\nu}{N_c}\right)\!M\,,\label{eq:munu}
\eeq
with $M$ the (constituent) mass of the quarks in the medium, we find that meson-like states are favored for $\smash{-M<\mu<\mu_\nu}$, while baryon-like states are preferred for $\smash{\mu_\nu<\mu<M}$. We give a simple explanation of these different regimes of $\mu$ while extending our findings and interpretations to probes in non-fundamental representations of the color group.

\begin{figure}[t]
    \centering
    \includegraphics[width=0.9\linewidth]{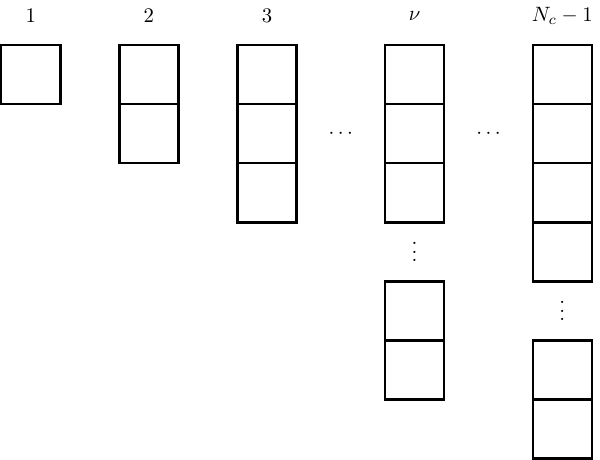}
   \caption{Young tableaux of the fundamental representations of SU($N_c$).}
    \label{fig:Young}
\end{figure}

Our analysis of the behavior of the net quark number gain is based on that of the Polyakov loop and the associated center symmetry \cite{Polyakov1978ThermalLiberation, Svetitsky1982CriticalTransitions}. Center symmetry is a subtle notion, however, which manifests itself naturally in the Euclidean formulation of QCD but has no obvious counterpart in the Minkowskian formulation where the Polyakov loop appears, instead, as a ratio of partition functions involving the QCD Hamiltonian. At the end of the paper, we present a rationale of how the main results of the present work could be recovered within the Minkowskian framework. The point is that one does not need the whole machinery of center symmetry, but, in a certain sense, only a remnant of it taking the form of a constraint on the states of the Hilbert space. More precisely, based on the center transformation properties of the Polyakov loop at finite quark chemical potential, we conjecture that the states with definite net quark number that enter the thermodynamical trace defining the Polyakov loop associated to an external probe in a given representation should themselves lie in a representation of $N_c$-ality opposite to that of the probe. As we also argue, these considerations bring some clarification to the counterintuitive observation that the center symmetry does not seem to follow the order/disorder paradigm of other broken symmetries, in the sense that the center-broken (resp. center-symmetric) phase corresponds to the high-temperature (resp. low-temperature) phase. 

\pagebreak

\section{General set-up}
We consider a bath of SU($N_c$) quarks and gluons at non-zero temperature $T$ and quark chemical potential $\mu$, and study the net quark number gained by the medium upon bringing a static color charge. 

\subsection{Fundamental representations}
We first consider the case where the probe lies in special types of irreducible representations of the color group, known as fundamental representations. Other irreducible representations (and also reducible ones) will be considered in Secs.~\ref{sec:other} and \ref{sec:tree}.

In the SU($N_c$) case, there exist $\smash{N_c-1}$ fundamental representations, obtained upon anti-symmetrizing tensor products of the (defining) representation associated with a quark. In representation theory, these correspond to single-column Young tableaux of size $\nu$, with $\smash{\nu=1,\dots,N_c-1}$ denoting the number of quarks involved, see Fig.~\ref{fig:Young}. Similarly, one can introduce anti-fundamental representations as anti-symmetrized tensor products of the representation associated with the antiquark. We denote them by $\bar\nu$, with $\smash{\nu=1,\dots,N_c-1}$ the number of antiquarks involved.

The fundamental representations $\nu$ and $N_c-\nu$, as well as the anti-fundamental representations $\bar\nu$ and $\overline{N_c-\nu}$, all have dimension 
\beq
C_{N_c}^\nu\equiv \frac{N_c!}{\nu!\,(N_c-\nu)!}\,.
\eeq
Of course, this does not necessarily mean that they are all equivalent, not even $\nu$ and $\bar\nu$, even though these two are related by charge conjugation. One result which is always true, however, is that $\bar\nu$ is equivalent to $N_c-\nu$. Thus, from the point of view of the color algebra, there is no way to distinguish an anti-symmetrized color state involving $\nu$ antiquarks from an anti-symmetrized color state involving $N_c-\nu$ quarks. This remark will be discussed further below when interpreting the results.

Because of the equivalence between fundamental and anti-fundamental representations, it is customary to list them as pairs $(\nu,\bar\nu)$ of charge-conjugated, fundamental and anti-fundamental representations, with $\smash{\nu=1,\dots,\lfloor N_c/2\rfloor}$. Generally, the two representations of a given pair are not equivalent, except when $N_c$ is even, in which case the representation $\smash{N_c/2}$ is equivalent to $\overline{N_c-N_c/2}=\overline{N_c/2}$. The two elements of the corresponding pair can then be considered as one and the same representation.

It is also customary to denote the irreducible representations using their dimension. Using boldface not to mix the two types of notation, we have thus various pairs of charge-conjugated fundamental representations ${\bf(C_{N_c}^\nu,\overline{C_{N_c}^\nu})}$, with $\smash{\nu=1,\dots,\lfloor N_c/2\rfloor}$ and where the last pair involves one and the same representation in the case where $N_c$ is even. In the SU(2) case, for instance, there is only one fundamental representation, ${\bf 2}$, which is equivalent to its charge-conjugated version ${\bf \bar 2}$. In the SU(3) case, there are two fundamental representations, ${\bf 3}$ and ${\bf \bar 3}$. In the SU(4) case, there are three fundamental representations ${\bf 4}$, ${\bf 6}$ and ${\bf \bar 4}$, with ${\bf 6}$ equivalent to ${\bf \bar 6}$.  

In what follows, we shall stick to the labeling of fundamental representations using the index $\nu$ whenever we derive general results, and, most of the time, we shall argue in terms of fundamental representations, with $\nu$ ranging from $1$ to $N_c-1$. The labeling in terms of dimensions will be used when considering illustrative examples for specific values of $N_c$.

\subsection{Net quark number}

To the fundamental representation $\nu$, one associates the Polyakov loop $\ell_\nu$ defined as the average
\beq
\ell_\nu\equiv\left\langle\frac{1}{C_{N_c}^\nu}{\rm tr}\,{\cal P}\,e^{ig\int_0^\beta d\tau\,A_4^a(\tau,\vec{x})t^a_\nu}\right\rangle,\label{eq:lnu}
\eeq
within Euclidean QCD at finite temperature $T$ and finite quark chemical potential $\mu$. The symbol ${\cal P}$ denotes time ordering along the Euclidean, compact time direction, $A_4^a$ is the temporal Euclidean gauge field, and the $t_\nu^a$ are the generators of the fundamental representation $\nu$. 

The Polyakov loop $\ell_\nu$ grants access to the free-energy $\Delta F_\nu$  gained by the bath upon bringing a charge in the considered representation \cite{Polyakov1978ThermalLiberation}\footnote{For a system at finite chemical potential, we have in mind the Landau free-energy, also known as grand potential. It is usually denoted $\Omega$, but here we keep the notation $F$ for simplicity.}
\beq
\Delta F_\nu=-T\ln \ell_\nu\,.
\eeq
Using the well-known relation $\smash{Q=-\partial F/\partial\mu}$ between the free-energy of the system and the average value of a conserved charge $Q$ of associated chemical potential $\mu$, we deduce that the net quark number response $\Delta Q_\nu$ of the bath upon bringing the probe is obtained as
\beq
\Delta Q_\nu=\frac{T}{\ell_\nu}\frac{\partial\ell_\nu}{\partial\mu}\,.\label{eq:Q}
\eeq
Since the fundamental representation $\smash{N_c-\nu}$ is equivalent to the anti-fundamental representation $\bar\nu$, which is itself the charge-conjugated of the fundamental representation $\nu$, one has\footnote{This involves two steps. First, one write $\smash{\ell_{N_c\nu}(\mu)=\ell_{\bar\nu}(\mu)}$, and then $\smash{\ell_{\bar\nu}(\mu)=\ell_\nu(-\mu)}$. Sometimes, $\ell_{\bar\nu}$ is also written as $\bar\ell_{\nu}$, but we shall refrain from doing so in this work.}
\beq
\ell_{N_c-\nu}(\mu)=\ell_\nu(-\mu)\,,
\eeq
and thus
\beq
\Delta Q_{N_c-\nu}(\mu)=-\Delta Q_\nu(-\mu)\,.\label{eq:C}
\eeq
In practice, this means that it is enough to study $\Delta Q_\nu$ or $\Delta Q_{N_c-\nu}$ for any $\mu$, or to study both $\Delta Q_\nu$ and $\Delta Q_{N_c-\nu}$ for $\mu\geq 0$. 

The net quark number response $\Delta Q_\nu$ is not the total net quark number gained by the medium because it does not include the quark number $\nu$ carried by the test charge in the fundamental representation $\nu$. The total quark number gained by the system is then
\beq
\Delta Q_\nu+\nu\,.\label{eq:total}
\eeq
Let us emphasize once more, however, that, from the point of view of the color algebra, the Polyakov loop $\ell_\nu$ in the fundamental representation stemming from $\nu$ quarks is not distinguishable from the Polyakov loop $\ell_{\overline{N_c-\nu}}$ in the anti-fundamental representation stemming from $N_c-\nu$ antiquarks. It follows that 
\beq
\Delta Q_{\nu}=\Delta Q_{\overline{N_c-\nu}}\,.\label{eq:equiv}
\eeq
Interpreting this as the net quark number response of the medium in the presence of a color probe in a fundamental representation originating from $\nu$ quarks, the net quark number gained by the system is indeed (\ref{eq:total}). However, interpreting (\ref{eq:equiv}) as the net quark number response of the medium in the presence of a color probe in an anti-fundamental representation originating from $N_c-\nu$ antiquarks, the net quark number gained by the system is
\beq
\Delta Q_{\overline{N_c-\nu}}-(N_c-\nu)=\Delta Q_\nu+\nu-N_c\,.
\eeq
 So, depending on the interpretation of the same color probe in terms of quarks or antiquarks, the net quark number gained by the system is shifted by $N_c$. This shift just reflects the fact that there exist distinct equivalent interpretations of the color probe in terms of quarks or antiquarks (as far as color is concerned). We thus expect our results to admit various equivalent and consistent interpretations, and we will verify that this is indeed the case.

To determine $\Delta Q_\nu$, we need access to the whole collection of fundamental Polyakov loops $\ell_\nu$. The latter are usually obtained from the extremization of the Polyakov loop potential $V(\ell_\nu)$. Interestingly enough, in the formal regime of QCD where all quarks are considered heavy, the behavior of $\ell_\nu$, and thus of $\Delta Q_\nu$, can be understood in terms of a few features, some of which are well-established properties of gauge theories at finite temperature. As discussed in Ref.~\cite{MariSurkau:2025how}, we believe that similar results should hold in QCD for temperatures significantly below the infrared constituent quark mass $\smash{T\ll M_q(p\to 0)\sim0.3\,}$GeV. There, the quark contribution to the Polyakov loop potential should be dominated by the infrared rainbow-resummed propagator whose constituent quark mass is determined by chiral symmetry breaking. We have tested this at the mean-field level in a Polyakov loop-extended NJL model, which effectively incorporates such an approximation \cite{MariSurkau:2025PNJLcsCF}.

\subsection{Heavy-quark regime}
In the heavy-quark regime, the Polyakov loop potential is well approximated by
\beq
V(\ell_\nu)=V_{\rm glue}(\ell_\nu)+V_{\rm quark}(\ell_\nu)\,,\label{eq: general V}
\eeq
with (see App.~\ref{app:pot} and Ref.~\cite{Reinosa:2019xqq})
\beq
& & V_{\rm quark}(\ell_\nu)=-\frac{TN_f}{\pi^2}\int_0^\infty dq\,q^2\nonumber\\
& & \hspace{0.2cm}\times\,\Bigg\{\ln \sum_{\nu=0}^{N_c} C_{N_c}^\nu \ell_\nu e^{-\nu\beta(\varepsilon_q-\mu)}\nonumber\\
& & \hspace{1.0cm}+\,\ln \sum_{\nu=0}^{N_c} C_{N_c}^\nu \ell_{N_c-\nu} e^{-\nu\beta(\varepsilon_q+\mu)}\Bigg\}\,,\label{eq:Vquark}
\eeq
where $\smash{\ell_0=\ell_{N_c}\equiv 1}$ and $\smash{\varepsilon_q\equiv\sqrt{q^2+M^2}}$. Higher loops involving quarks are suppressed due to the large quark masses. For simplicity, we consider $N_f$ degenerate quarks of mass $M$, but the discussion can be extended to non-degenerate quarks. 

As for the glue contribution $V_{\rm glue}(\ell_\nu)$, we know that it is center-symmetric \cite{Svetitsky1982CriticalTransitions} 
\beq
V_{\rm glue}(\ell_\nu)=V_{\rm glue}(e^{-i2\pi \nu/N_c}\ell_\nu)\,,\label{eq:center}
\eeq
and confining at low temperatures, that is, its relevant extremum in this limit is located at $\smash{\ell_\nu=0}$, as confirmed by lattice simulations \cite{McLerran:1981pb, Brown:1988qe, Gupta:2007ax, Lucini:2012gg, Lo:2013hla}. In line with what happens in the SU(3) case \cite{MariSurkau:2025how}, we shall assume that, for $\smash{|\mu|<M}$, the glue contribution dominates over the exponentially suppressed quark contribution at low temperatures. Strictly speaking, this is not a rigorously established fact but rather the phenomenological observation that most, if not all, continuum approaches that successfully capture the confinement/deconfinement transition in the glue sector share this property \cite{Lo:2013hla, Fukushima2004ChiralLoop, Ratti:2005jh, Roessner:2006xn, Fukushima:2008wg, Reinosa:2014ooa, MariavanEgmond2022ATemperature,  Braun:2007bx}. Numerical tests (through lattice Monte-Carlo simulations with heavy quarks \cite{Fromm2012ThePotentials}) of the results of this work may shed some light on the validity of this assumption. For $\smash{|\mu|\geq M}$, the quark contribution is only polynomially suppressed, and the discussion is more subtle, see Ref.~\cite{MariSurkau:2025how}.

\section{Low-temperature limit}
The equations determining the various $\ell_\nu$ found by extremizing the potential (\ref{eq: general V}) write
\begin{eqnarray}
\frac{\partial V_{\rm glue}}{\partial\ell_{N_c-\nu}} & = & \frac{TN_f}{\pi^2}\int_0^\infty dq\,q^2\nonumber\\
& \times & \Bigg\{\frac{C_{N_c}^\nu e^{-(N_c-\nu)\beta(\varepsilon_q-\mu)}}{\sum_{\nu=0}^{N_c} C_{N_c}^\nu \ell_\nu e^{-\nu\beta(\varepsilon_q-\mu)}}\nonumber\\
& & \hspace{0.1cm}+\,\frac{C_{N_c}^\nu e^{-\nu\beta(\varepsilon_q+\mu)}}{\sum_{\nu=0}^{N_c} C_{N_c}^\nu \ell_{N_c-\nu} e^{-\nu\beta(\varepsilon_q+\mu)}}\Bigg\}\,,\label{eq:gg1}
\end{eqnarray}
where $\smash{\nu=1,\dots, N_c-1}$ and we have chosen to take a derivative with respect to $\ell_{N_c-\nu}$ for later convenience. Let us now argue that the behavior of $\ell_\nu$ and thus of $\Delta Q_\nu+\nu$ at low temperatures can be inferred from the few ingredients given above. We shall focus on the case $\smash{|\mu|<M}$ and briefly comment on the case $\smash{|\mu|\geq M}$, which can be dealt with following Ref.~\cite{MariSurkau:2025how}.

\subsection{Asymptotic behavior}
Assuming that $\smash{|\mu|<M}$, any exponential $e^{-\nu\beta(\varepsilon_q\pm\mu)}$ appearing in the right-hand side of Eq.~(\ref{eq:gg1}), is bounded by
\beq
e^{-\nu\beta(\varepsilon_q\pm\mu)}\leq e^{-\nu\beta(M-|\mu|)}\,,
\eeq
which is independent of the integration variable $q$ and which goes to $0$ exponentially as $\smash{T\to 0}$. This means that, in this case, we can safely approximate the above equations as
\begin{eqnarray}
\frac{\partial V_{\rm glue}}{\partial\ell_{N_c-\nu}} & = & \frac{C_{N_c}^\nu TN_f}{\pi^2}\\
& \times & \int_0^\infty dq\,q^2\Big(e^{-(N_c-\nu)\beta(\varepsilon_q-\mu)}+\,e^{-\nu\beta(\varepsilon_q+\mu)}\Big)\,,\nonumber\label{eq:g3}
\end{eqnarray}
where we note that the quark contribution no longer depends on the Polyakov loops. Moreover, the $\mu$-dependence can be explicitly pulled out of the integrals. More precisely, upon introducing the function
\beq
f_y\equiv \frac{1}{\pi^2}\int_0^\infty \!\!\! dx\,x^2e^{-y\sqrt{x^2+1}}\sim \frac{y^{-3/2}}{\sqrt{2}\pi^{3/2}}e^{-y}\,,\label{eq:f}
\eeq
where the asymptotic expression on the right-hand side is understood in the limit $\smash{y\to\infty}$, the equations become
\beq
\frac{\partial V_{\rm glue}}{\partial\ell_{N_c-\nu}} & \simeq & \alpha_\nu\Big(e^{(N_c-\nu)\beta\mu} f_{(N_c-\nu)\beta M}+e^{-\nu\beta\mu} f_{\nu\beta M}\Big)\,,\nonumber\\\label{eq:gap}
\eeq
with $\smash{\alpha_\nu\equiv C_{N_c}^\nu N_f TM^3}$ and $\smash{\nu=1,\dots,N_c-1}$.

Since $\smash{|\mu|<M}$, the right-hand side of Eq.~(\ref{eq:gap}) approaches $0$ exponentially as $\smash{T\to 0}$, and, because we have assumed that the glue potential does not vanish so rapidly in comparison and is confining at low temperatures, we deduce that the $\ell_\nu$ approach $0$ in that limit. The set of equations (\ref{eq:gap}) can then be linearized around $\smash{\ell_\nu=0}$:
\beq
& & \frac{\partial^2V_{\rm glue}}{\partial\ell_{N_c-\nu}\partial\ell_{{\nu'}}}\ell_{{\nu'}}\label{eq:gap3}\\
& & \hspace{0.7cm}\simeq\,\alpha_\nu\Big(e^{(N_c-\nu)\beta\mu} f_{(N_c-\nu)\beta M}+e^{-\nu\beta\mu} f_{\nu\beta M}\Big)\,,\nonumber
\eeq
where a summation over $\nu'$ is implied and the Hessian $\partial^2V_{\rm glue}/\partial\ell_{N_c-\nu}\partial\ell_{\nu'}$ is evaluated at the confining, or center-symmetric, point where all $\ell_\nu$ vanish. Now, from Eq.~(\ref{eq:center}), all derivatives $\partial^2V_{\rm glue}/\partial\ell_{N_c-\nu}\partial\ell_{\nu'}$ vanish at the confining point, except when $N_c-\nu+\nu'$ is a multiple of $N_c$. But because $N_c-\nu$ and ${\nu'}$ are comprised between $1$ and $N_c-1$, this only occurs when $\smash{N_c-\nu+\nu'=N_c}$, that is when $\smash{\nu=\nu'}$. It follows that
\beq
\ell_{\nu}\simeq\frac{\alpha_\nu}{X_\nu}\Big(e^{(N_c-\nu)\beta\mu} f_{(N_c-\nu)\beta M}+e^{-\nu\beta\mu} f_{\nu\beta M}\Big)\,,\label{eq:gap0}
\eeq
where $\smash{X_\nu\equiv\partial^2V_{\rm glue}}/\partial\ell_{N_c-\nu}\partial\ell_\nu$ is evaluated at the center-symmetric point. Taking a $\mu$-derivative, this gives
\beq
\frac{\partial\ell_{\nu}}{\partial\mu} & \simeq & \beta\frac{\alpha_\nu}{X_\nu}\Big((N_c-\nu)e^{(N_c-\nu)\beta\mu} f_{(N_c-\nu)\beta M}\nonumber\\
& & \hspace{1.5cm}-\,\nu e^{-\nu\beta\mu} f_{\nu\beta M}\Big)\,,
\eeq
which, combined with Eq.~(\ref{eq:gap0}) as in Eq.~(\ref{eq:Q}), yields
\beq
\Delta Q_\nu & \simeq & \frac{(N_c-\nu)e^{(N_c-\nu)\beta\mu} f_{(N_c-\nu)\beta M}-\nu e^{-\nu\beta\mu} f_{\nu\beta M}}{e^{(N_c-\nu)\beta\mu} f_{(N_c-\nu)\beta M}+e^{-\nu\beta\mu} f_{\nu\beta M}}\,,\nonumber\\
\eeq
which verifies (\ref{eq:C}) as expected and which we rewrite as
\beq
\Delta Q_\nu\simeq\frac{(N_c-\nu)-\nu e^{-N_c\beta\mu} f_{\nu\beta M}/f_{(N_c-\nu)\beta M}}{1+e^{-N_c\beta\mu} f_{\nu\beta M}/f_{(N_c-\nu)\beta M}}\,.
\eeq
Using Eq.~(\ref{eq:f}), we find
\beq
\Delta Q_\nu & \simeq & \frac{(N_c-\nu)-c\,\nu\,e^{-N_c\beta(\mu-\mu_\nu)}}{1+c\,e^{-N_c\beta(\mu-\mu_\nu)}}\,,
\eeq
with $\mu_\nu$ defined in Eq.~(\ref{eq:munu}) and $c$ a constant that depends on $N_c$ and $\nu$.  

\begin{figure}[t]
    \centering
    \includegraphics[width=0.9\linewidth]{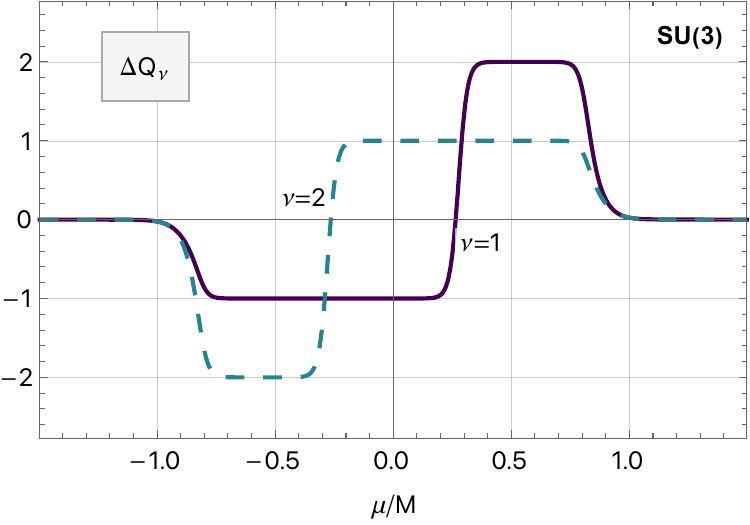}
    \caption{Net quark number responses $\Delta Q_{1,2}$ of a thermal bath upon bringing one- or two-quark probes in a fundamental representation of SU(3), at low temperature, and as functions of $\mu/M$.}
    \label{fig: su3 fund DQ}
\end{figure}

It follows from these formulas that the net quark number response of the bath approaches $-\nu$ or $N_c-\nu$ at low temperatures, depending on whether $\smash{\mu<\mu_\nu}$ or $\smash{\mu>\mu_\nu}$ (and $\smash{|\mu|<M}$). In other words, the net quark number response of the bath becomes a step function jumping from $-\nu$ to $N_c-\nu$ at this particular value of $\mu$:
\beq
\Delta Q_\nu=\left\{
\begin{array}{cl}
-\nu\,, & \mbox{for } \mu<\mu_\nu\,,\\
N_c-\nu\,, & \mbox{for } \mu>\mu_\nu\,,
\end{array}\label{eq:response}
\right.
\eeq
see Fig.~\ref{fig: su3 fund DQ} for an illustration in the SU(3) case. 

\begin{figure}[t]
    \centering
    \includegraphics[width=0.9\linewidth]{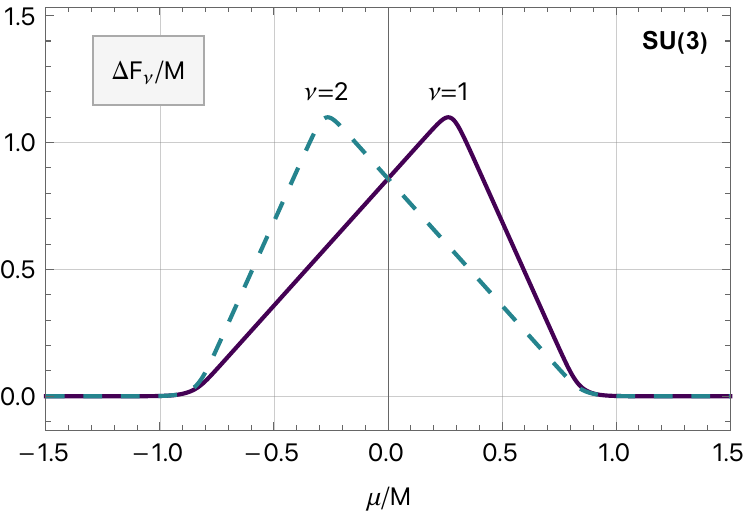}
    \caption{Free energies $\Delta F_{1,2}$ gained by a thermal bath upon bringing one- or two-quark probes in a fundamental representation of SU(3) scaled by the quark mass $M$, at low temperature, and as functions of $\mu/M$.}
    \label{fig: su3 fund log L}
\end{figure}

This zero-temperature limit in the range $\smash{|\mu|<M}$ is rooted in the corresponding low temperature behavior (\ref{eq:gap0}) of the Polyakov loops which we will retain as
\beq
\frac{\ell_\nu}{(\beta M)^{-3/2}}\propto\left\{
\begin{array}{cl}
e^{-\nu\beta(\mu+M)}\,, & \mbox{for } \mu<\mu_\nu\,,\\
e^{(N_c-\nu)\beta(\mu-M)}\,, & \mbox{for } \mu>\mu_\nu\,.
\end{array}\label{eq:fund}
\right.
\eeq
Thus, what matters is not how small $\ell_\nu$ is but how its suppression depends on $\mu$. In the confining phase, the suppression is exponential with an argument that varies linearly over certain ranges of $\mu$. This is nicely visualized if, instead of plotting $\ell_\nu$, one represents its logarithm, see e.g. \cite{MariSurkau:2025xqcd}. With an extra factor of $-T$, this is nothing but the free-energy difference $\smash{\Delta F_\nu=-T\ln\ell_\nu}$:
\beq
\Delta F_\nu=\left\{
\begin{array}{cl}
\nu(\mu+M)\,, & \mbox{for } \mu<\mu_\nu\,,\\
(N_c-\nu)(M-\mu)\,, & \mbox{for } \mu>\mu_\nu\,.\\
\end{array}
\right.
\eeq
see Fig.~\ref{fig: su3 fund log L} for an illustration in the SU(3) case.

We emphasize that these results were obtained using a linear approximation, as shown in Eq.~(\ref{eq:gap3}). The latter, however, is not totally accurate for $\beta M$ large but finite. This is because, over certain ranges of $\mu$, higher terms in the Taylor expansion of $\partial V_{\rm glue}/\partial\ell_{N_c-\nu}$ can have the same exponential suppression as the linear terms when the asymptotic expressions (\ref{eq:gap0}) are assumed. Yet, the linear approximation becomes exact for infinitely large $\beta M$ due to additional suppressions by powers of $\beta M$. Two key ingredients in proving this result are\footnote{Rather than a rigorous proof, this is more of a consistency check.} 1) center symmetry, which implies that all terms in the Taylor expansion of $\partial V_{\rm glue}/\partial\ell_{N_c-\nu}$,
\beq
\frac{\partial V_{\rm glue}}{\partial\ell_{N_c-\nu}}=\sum_{p\geq 1}\frac{1}{p!}\frac{\partial^{p+1} V_{\rm glue}}{\partial\ell_{N_c-\nu}\partial\ell_{\nu_1}\cdots\partial\ell_{\nu_p}}\ell_{\nu_1}\cdots\ell_{\nu_p}\,,
\eeq
 are constrained by
 \beq
 \nu_1+\cdots+\nu_p=\nu+N_c k\,,\label{eq:29}
 \eeq
 for some positive integer $k$, and 2) that,\footnote{As discussed in App.~\ref{app:ordering}, this result applies also to the case where some of the $\nu_i$ or $\nu$ are equal to $0$, representing the trivial representation, in which case the corresponding Polyakov loops are equal to $1$.} for any $\smash{p\geq 2}$,
 \beq
 \frac{\ell_{\nu_1}\cdots\ell_{\nu_p}}{\ell_\nu}\to 0\,,\label{eq:30}
 \eeq
 as $\smash{T\to 0}$ for any $\smash{|\mu|<M}$. The proof of (\ref{eq:30}) is a bit technical, so we relegate it to App.~\ref{app:ordering}, together with a strategy to go beyond the linear approximation when $\beta M$ is large but not infinitely large in App.~\ref{app:linear}. The result (\ref{eq:30}) will also play a role when discussing non-fundamental representations in Sec.~\ref{sec:other}.

\subsection{Interpretation}
\begin{figure}
    \centering
    \includegraphics[width=0.9\linewidth]{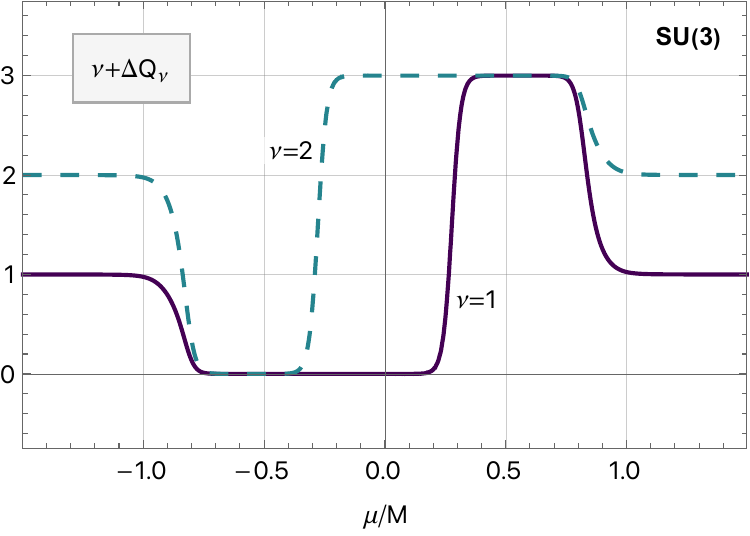}
    \caption{Net quark numbers $1+\Delta Q_1$ and $2+\Delta Q_2$ gained by a thermal bath upon bringing one- or two-quark probes in a fundamental representation of SU(3), at low temperature, and as functions of $\mu/M$.}
    \label{fig:illustration_su3}
\end{figure}

The low temperature behavior (\ref{eq:response}) for the net quark number response of the bath implies that the net quark number gained by the system is
\beq
\Delta Q_\nu+\nu=\left\{
\begin{array}{cl}
0\,, & \mbox{for } \mu<\mu_\nu\,,\\
N_c\,, & \mbox{for } \mu>\mu_\nu\,,
\end{array}
\right.
\eeq
see Fig.~\ref{fig:illustration_su3} for an illustration in the SU(3) case. We can interpret this result as follows. 

For very negative $\mu$, there is a large excess of antiquarks w.r.t. to quarks in the medium, so it is simpler for the $\nu$ quarks of the probe to combine with $\nu$ antiquarks of the medium ($\smash{\Delta Q_\nu=-\nu}$), forming $\nu$ mesons that absorb the color probe, hence $\smash{\Delta Q_\nu+\nu=0}$, see Fig.~\ref{fig:illustration_su3}. In contrast, for very positive $\mu$, there is a large excess of quarks w.r.t to antiquarks and then it is simpler for the $\nu$ quarks of the probe to combine with $N_c-\nu$ quarks of the medium ($\smash{\Delta Q_\nu=N_c-\nu}$), forming a baryon to absorb the color probe, hence $\smash{\Delta Q_\nu+\nu=N_c}$, see Fig.~\ref{fig:illustration_su3}.

We also note that, for $\smash{\mu=0}$, in the absence of a probe, there is the same number of quarks and antiquarks in the medium. When bringing a probe made of $\nu$ quarks into such a medium, it can combine either with $\nu$ antiquarks or with $N_c-\nu$ quarks of the medium, but it is always simpler to combine with the least number of particles. Then, when $\smash{\nu< N_c-\nu}$, that is when $\smash{\nu< N_c/2}$, $\nu$ mesons are formed, while for $\smash{\nu> N_c/2}$, a baryon is formed. This is also clearly visible in Fig.~\ref{fig:illustration_su3}. It implies that the boundary value of $\mu$ between the two regimes has to vanish formally for $\smash{\nu=N_c/2}$, which is what we find in Eq.~(\ref{eq:munu}). Of course, since $\nu$ is an integer, this can only happen in practice in the case where $N_c$ is even.

In fact, the boundary value $\mu_\nu$ could have been anticipated as follows. In the low-temperature limit, the partition function is dominated by states that minimize $\smash{H-\mu\,Q}$. Let us then compare this quantity as the medium brings out either $\nu$ antiquarks or $N_c-\nu$ quarks, assuming that their energy is mostly given by the large mass $M$. In the first case
\beq
H-\mu\,Q\simeq \nu M-\mu(-\nu)=\nu(M+\mu)\,,\label{eq:one}
\eeq
while in the second case
\beq
H-\mu\,Q & \simeq & (N_c-\nu)M-\mu(N_c-\nu)\nonumber\\
& = & (N_c-\nu)(M-\mu)\,.\label{eq:two}
\eeq
The right-hand side of Eq.~(\ref{eq:one}) is obviously smaller than the right-hand side of Eq.~(\ref{eq:two}) for very negative $\mu$, corresponding to the system bringing out $\nu$ antiquarks, while it is the opposite for very positive $\mu$, corresponding to the system bringing out $N_c-\nu$ quarks. The value of $\mu$ at which the transition occurs is obtained by equating the two right-hand sides:
\beq
\nu(M+\mu)=(N_c-\nu)(M-\mu)\,,
\eeq
which leads to the boundary value in Eq.~(\ref{eq:munu}). Of course, our calculation does not give access to the actual color representation of the states formed to absorb the charge, so we should be more careful when referring to these states as mesons or baryons. What we can say is that the states formed to screen the probes have the same net quark number as mesons or baryons. In Sec.~\ref{sec:other}, we will discuss in which form this result generalizes to other (non-fundamental) representations and in Sec.~\ref{sec:color}, we investigate its relation with color.

\begin{figure}[t]
    \centering
    \includegraphics[width=0.9\linewidth]{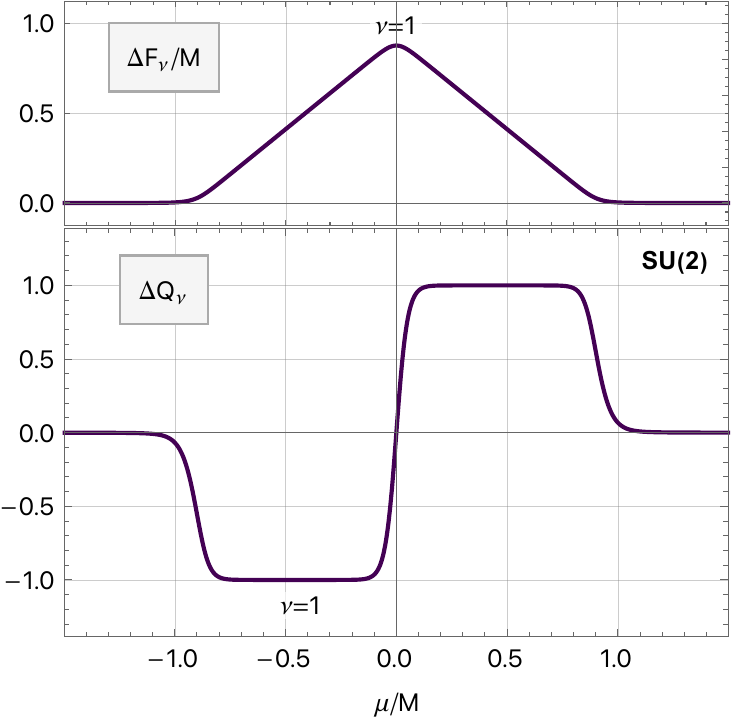}
    \caption{Top: Free energy $\Delta F_1$ gained by a thermal bath upon bringing a quark probe into the fundamental representation of SU(2) scaled by the quark mass $M$, at low temperature, and as a function of $\mu/M$. Bottom: Corresponding net quark number responses.}
    \label{fig: su2 fund DQ}
\end{figure}

Recall also that, from the point of view of color, a probe in the fundamental representation $\nu$, corresponding to $\nu$ quarks, is indistinguishable from a probe in the anti-fundamental representation $\overline{N_c-\nu}$, corresponding to $N_c-\nu$ antiquarks. As we have seen above, with this interpretation, the total net quark number gained by the medium is shifted by $-N_c$. It is essentially equal to $-N_c$ or $0$ depending on whether $\smash{\mu<\mu_\nu}$ or $\smash{\mu>\mu_\nu}$. The interpretation of this result, despite being different than before, is still consistent. Indeed, for very negative $\mu$, there is a larger number of antiquarks in the bath, and so it is simpler for the $N_c-\nu$ antiquarks of the probe to combine with $\nu$ antiquarks of the bath ($\smash{\Delta Q_{\overline{N_c-\nu}}=-\nu}$) to form one anti-baryon, hence $\smash{\Delta Q_{\overline{N_c-\nu}}-(N_c-\nu)=-N_c}$. On the contrary, for very positive $\mu$, there is a larger number of quarks in the bath and so it is simpler for the $N_c-\nu$ antiquarks of the probe to combine with $N_c-\nu$ quarks ($\smash{\Delta Q_{\overline{N_c-\nu}}=N_c-\mu}$) to form $N_c-\nu$ mesons, hence $\smash{\Delta Q_{\overline{N_c-\nu}}-(N_c-\nu)=0}$. In what follows, to accommodate more easily these different interpretations, we will plot directly the quantities $\smash{\Delta Q_\nu}$, as in Fig.~\ref{fig: su3 fund DQ}.

\begin{figure}[t]
    \centering
    \includegraphics[width=0.9\linewidth]{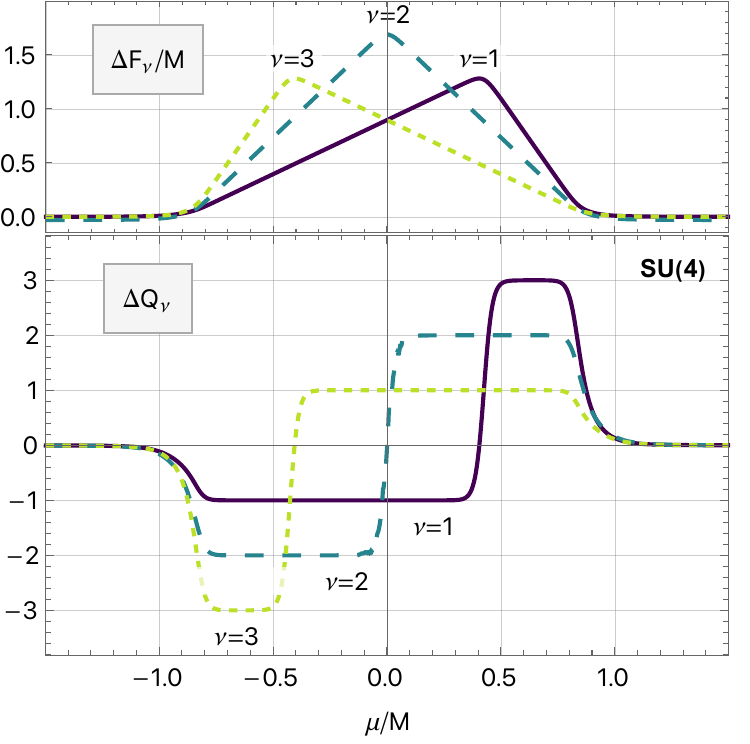}
    \caption{Top: Free energies $\Delta F_{1,2,3}$ gained by a thermal bath upon bringing one-, two- or three-quark probles in a fundamental representation of SU(4) scaled by the quark mass $M$, at low temperature, and as functions of $\mu/M$. Bottom: Corresponding net quark number responses.}
    \label{fig: su4 fund DQ}
\end{figure}

We stress that the above results have been derived using just a few assumptions regarding the glue potential. We have verified these results numerically using various models for the glue potential that comply with these assumptions. In particular, the plots shown in the present paper use the model of Ref.~\cite{MariavanEgmond2022ATemperature}, see Refs.~\cite{MariSurkau2024DeconfinementDependences, MariSurkau2025Heavy-quarkModel} for details on evaluation for a general $N_c$, and at finite, real chemical potentials. Within this model, we confirm that our predictions apply not only to the SU(3) case but also to SU(2) or SU(4), see Figs.~\ref{fig: su2 fund DQ} and \ref{fig: su4 fund DQ}. For SU(3), various popular phenomenological model potentials are available in the literature, and for SU(2) and SU(4), we compared with the model of \cite{Reinosa:2014ooa}. In all cases, we found the results to be model-independent. We also give some details on how the Polyakov loop potentials have to be calculated at finite, real $\mu$ for $N_c\geq3$ in App.~\ref{app:real_mu}, to ensure real-valued Polyakov loops, avoiding a mild, continuum sign problem \cite{Reinosa:2019xqq}.

Finally, let us briefly comment on what happens when $\smash{|\mu|\geq M}$. In this case, the quark contribution to the Polyakov loop potential is only suppressed polynomially, and the argumentation depends more crucially on the properties of the glue potential at low temperatures. In Ref.~\cite{MariSurkau:2025how}, however, we have argued that the behavior of $\Delta Q_1$ and $\Delta Q_2$ in the SU(3) case is again model independent and that both quantities become negligible as $\smash{T\to 0}$ for $|\mu|\geq M$, signaling a deconfined phase.\footnote{As for the Polyakov loops, we have found that their behavior as $\smash{T\to 0}$ and $\smash{|\mu|\geq M}$ can depend on the considered model. For instance, the Polyakov loops may approach $1$ or a value close to $1$, but in some cases they can also approach $0$. Even in this case, however, the region $\smash{|\mu|>M}$ should be considered as a deconfined phase in the sense that the Polyakov loops approach $0$ way more slowly than what happens for $\smash{|\mu|<M}$. An even better way to argue is that, independently of the model, the free-energy gains approach $0$ for $\smash{|\mu|>M}$ while they converge to finite non-zero values for $\smash{|\mu|<M}$.} A similar reasoning can be applied to the SU($N_c$) case and, in the examples we have considered, we find indeed that $\Delta Q_\nu$ is strongly suppressed for $\smash{|\mu|>M}$, see Figs.~\ref{fig: su3 fund DQ}, \ref{fig: su2 fund DQ} and \ref{fig: su4 fund DQ}.

\subsection{Ordering}
The above results allow us to compare the energy costs for bringing a bunch of $\nu_1$ quarks or a bunch of $\smash{\nu_2>\nu_1}$ quarks into the bath, assuming that each bunch comes in a fundamental representation (other representations will be considered in Secs.~\ref{sec:other} and \ref{sec:tree}). To this end, we estimate the magnitude of $\ell_{\nu_1}/\ell_{\nu_2}$ at low temperatures and for $\mu$ in the range $\smash{|\mu|<M}$. There are three intervals of $\mu$ to be considered.

For $\smash{\mu_{\nu_1}<\mu<M}$, using Eq.~(\ref{eq:fund}), we find
\beq
\frac{\ell_{\nu_1}}{\ell_{\nu_2}}\propto e^{(\nu_2-\nu_1)\beta(\mu-M)}\to 0\,,\label{eq:lim1}
\eeq
whereas for $\smash{-M<\mu<\mu_{\nu_2}}$\,, we find
\beq
\frac{\ell_{\nu_1}}{\ell_{\nu_2}}\propto e^{(\nu_2-\nu_1)\beta(\mu+M)}\to \infty\,.\label{eq:lim2}
\eeq
Finally, in the intermediate interval $\smash{\mu_{\nu_2}<\mu<\mu_{\nu_1}}$, we find
\beq
\frac{\ell_{\nu_1}}{\ell_{\nu_2}}\propto e^{-(N_c+\nu_1-\nu_2)\beta\mu+(N_c-\nu_1-\nu_2)\beta M}\,.
\eeq
The transition between the two types of limits (\ref{eq:lim1}) and (\ref{eq:lim2}) occurs for
\beq
\mu_{\nu_1,\nu_2}=\frac{N_c-\nu_1-\nu_2}{N_c+\nu_1-\nu_2}\,M\,,\label{eq:mu12}
\eeq
which lies necessarily between $\mu_{\nu_1}$ and $\mu_{\nu_2}$. This can be explicitly checked by noticing that $\mu_{\nu_1,\nu_2}$ rewrites as
\beq
\mu_{\nu_1,\nu_2}=(1-\alpha)\mu_{\nu_1}+\alpha\,\mu_{\nu_2}\,,
\eeq
with
\beq
\alpha=\frac{\nu_1}{N_c-\nu_2+\nu_1}>0\,.
\eeq
and is then located within the considered interval.

By construction, $\mu_{\nu_1,\nu_2}$ is the value of $\mu$ at which $\smash{\Delta F_{\nu_1}=\Delta F_{\mu_2}}$. In fact, $\smash{\Delta F_{\nu_1}<\Delta F_{\nu_2}}$ for $\smash{\mu<\mu_{\nu_1,\nu_2}}$, and $\smash{\Delta F_{\nu_1}>\Delta F_{\nu_2}}$ for $\smash{\mu>\mu_{\nu_1,\nu_2}}$, see for instance Fig.~\ref{fig: su4 fund DQ}. This makes sense physically. For very negative $\mu$, there is a large excess of antiquarks over quarks in the bath, and we know that the $\nu_1$ quarks will combine with $\nu_1$ antiquarks to form $\nu_1$ mesons, whereas the $\nu_2$ quarks will combine with $\nu_2$ antiquarks to form $\nu_2$ mesons. Since $\nu_1<\nu_2$, it will be simpler to form the $\nu_1$ mesons than the $\nu_2$ mesons. In contrast, for very positive $\mu$, there is a large excess of quarks over antiquarks in the bath, and we know that the $\nu_1$ quarks will combine with $N_c-\nu_1$ quarks to form one baryon, whereas the $\nu_2$ quarks will combine with $N_c-\nu_2$ quarks to form one baryon. Since $\smash{\nu_1<\nu_2}$, and thus $\smash{N_c-\nu_2<N_c-\nu_1}$, it will be simpler to form the latter baryon.

As before, we note that, for $\smash{\mu=0}$, in the absence of a probe, there is the same number of quarks and antiquarks in the medium. When bringing a probe made of $\nu_i$ quarks in a fundamental representation, it can combine either with $\nu_i$ antiquarks or with $N_c-\nu_i$ quarks of the medium, but it is always simpler to combine with the least number of particles, that is, with ${\rm Min}\,(\nu_i,N_c-\nu_i)$. So for bringing $\nu_1$ quarks to be  energetically equivalent to bringing $\nu_2$ quarks into the bath at $\smash{\mu=0}$, we should have either $\smash{\nu_1=\nu_2}$ or $\smash{\nu_1=N_c-\nu_2}$. Since we have assumed that $\smash{\nu_1\neq\nu_2}$, we deduce that $\mu_{\nu_1,\nu_2}$ should vanish for $\smash{\nu_1+\nu_2=N_c}$, which is what we find in Eq.~(\ref{eq:mu12}).

\section{High-temperature limit}

We saw that $\Delta Q_\nu$ is suppressed at low temperatures and $\smash{|\mu|\geq M}$. Another limit where we should expect $\Delta Q_\nu$ to be strongly suppressed is the high-temperature limit $\smash{T\to\infty}$. Indeed, the deconfined phase being dominated by quark degrees of freedom, the addition of one quark/antiquark should increase/decrease the net quark number by one unit only. Here, however, we cannot rely on Eq.~(\ref{eq:Vquark}) simply because it is not valid for $\smash{T\gg M}$. Instead, this regime can be studied following the techniques in Ref.~\cite{Dumitru:2013xna} and the related function
\beq
W(r^j)=V(\ell_\nu(r^j))\,,
\eeq
with\footnote{Our notation differs slightly from the one in Ref.~\cite{Dumitru:2013xna}.}
\beq
\ell_\nu(r^j)=\frac{1}{C_{N_c}^\nu}\sum_{\rho_\nu} e^{i\rho_\nu^j r^j}\,,
\eeq
where $r$ denotes a vector with $N_c-1$ components and $\rho_\nu$ runs over the weights of the fundamental representation $\nu$, see App.~\ref{app:pot}, which are also vectors with $N_c-1$ components. Note that the weights are always real, but this is not necessarily the case for $r$. In fact, this depends on the chemical potential. In the SU(3) case, for instance, when the chemical potential is real, while the first component of $r$ is taken real, the second component should be taken purely imaginary, see Ref.~\cite{Reinosa:2019xqq}. The extension to the SU($N_c$) case is discussed in App.~\ref{app:real_mu} and was taken into account when producing the SU(4) results above.

Returning to the question of the high-temperature limit, we can use an argument based on dimensional analysis and the renormalization group. Denoting the renormalization scale by $s$, we can indeed write
\beq
& & W(r^j;T,\mu;s,g(s),M(s))\nonumber\\
& & \hspace{1.0cm}=\,W(r^j;T,\mu;T,g(T),M(T))\nonumber\\
& & \hspace{1.0cm}=\,T^4W(r^j;1,\mu/T;1,g(T),M(T)/T)\,.
\eeq
Thus, for large temperatures, one can use perturbation theory, and, at leading order, one is led to consider the Weiss potential
\beq
& & W(r^j;T,\mu;s_0,g(s_0),M(s_0))\nonumber\\
& & \hspace{0.7cm}\simeq\,T^4w_{\rm Weiss}(r^j;\mu/T)\nonumber\\
& & \hspace{0.7cm}\simeq\,T^4\left[w_{\rm Weiss}(r^j;0)+\frac{\mu}{T}w^{(0,\dots,0,1)}_{\rm Weiss}(r^j;0)\right]\!.
\eeq
\noindent{Since the extremum of $w_{\rm Weiss}(r^j;0)$ is located at $\smash{\{r^j=0\}}$, we deduce that the extremum $r^j(T)$ of the complete potential obeys}
\beq
r^j(T)=a^j(T)+\frac{\mu}{T}b^j(T)+{\cal O}\left(\frac{\mu^2}{T^2}\right)\,,
\eeq
with $\smash{a^j(T)\to 0}$. The Polyakov loop is then
\beq
\ell_\nu= \ell_\nu(a)+ \frac{\mu}{T}\frac{b^j}{N_c}\sum_{\rho_\nu} \rho_\nu^j\,e^{i\rho_\nu^j a^j}+{\cal O}\left(\frac{\mu^2}{T^2}\right)\,.
\eeq
It follows that
\beq
\Delta Q_\nu\simeq\frac{\frac{b^k}{N_c}\sum_{\rho_\nu} \rho_\nu^k\,e^{\pm i\rho_\nu^j a^j}}{1+\frac{\mu}{T}\frac{b^k}{N_c}\sum_{\rho_\nu} \rho_\nu^k\,e^{\pm i\rho_\nu^k a^k}}\,,
\eeq
which approaches $0$ as $\smash{T\to\infty}$ since $\smash{a_j\to 0}$ and $\smash{\sum_{\rho_\nu} \rho_\nu^j=0}$, see App.~\ref{app:pot}.

In the heavy-quark case, following Ref.~\cite{MariSurkau:2025how}, it can be argued that this asymptotic result extends to most of the deconfined phase above the transition line $T_c(\mu)$. Similarly, most of the region below this line displays the step-function $\Delta Q_\nu$ behavior discussed above, with a $T$-dependent transition value $\mu_\nu(T)$ whose determination requires going beyond the linear approximation used in Eq.~(\ref{eq:gap3}), see App.~\ref{app:linear}.

\section{Non-fundamental representations}\label{sec:other}

The fundamental representations that we have considered so far are a special type of irreducible representations corresponding to single-column Young tableaux, see Fig.~\ref{fig:Young}. Here, we would like to investigate irreducible representations corresponding to Young tableaux with more than one column,\footnote{Recall that the size of the columns of a Young tableau cannot exceed $N_c-1$ and should decrease from left to right. The size of the lines should also decrease from top to bottom but it is not limited in number.} as well as reducible representations obtained as tensor products or direct sums of irreducible representations.

To each such representation $R$ of dimension $d_R$, one can associate a Polyakov loop
\beq
\ell_R\equiv\left\langle\frac{1}{d_R}{\rm tr}\,{\cal P}\,e^{ig\int_0^\beta d\tau\,A_4^a(\tau,\vec{x})t^a_R}\right\rangle.\label{eq:lR}
\eeq
By analogy with the case of the fundamental representations, we will assume that
\beq
\Delta F_R=-T\ln\ell_R\label{eq:FR}
\eeq
corresponds to the free-energy gained by the bath upon bringing a probe in the color representation $R$, and then, that
\beq
\Delta Q_R=\frac{T}{\ell_R}\frac{\partial\ell_R}{\partial\mu}
\eeq
gives the net quark number response of the bath upon bringing that probe. We stress that these interpretations are far from obvious\footnote{It can be shown that all color states of a probe in a given fundamental representation can be connected to each other via appropriate color transformations. This implies that the energy cost for bringing such a probe into the thermal bath does not depend on the particular color state of the probe in the chosen representation. This is far from obvious for a probe in a non-fundamental representation since it is not true anymore that all the states of the representation can be connected with each other via appropriate color transformations.} but we would like to investigate how far one can go with them. A related aspect, which applies to both fundamental and non-fundamental representations, is that it is not obvious why the average in the right-hand side of (\ref{eq:lR}) is positive, as required by the free-energy interpretation. This is not obvious even at vanishing chemical potential, where the quark determinant is positive.

\subsection{$N_c$-ality of SU($N_c$) representations}

A particularly important notion in what follows will be that of the $N_c$-ality of an SU($N_c$) representation, which we now recall by first looking at the case of an irreducible representation $R$. There is one and only one Young tableau ${\cal T}_R$ attached to $R$. The $N_c$-ality $\nu_R$ of $R$ is then obtained by counting the number of elements of ${\cal T}_R$ while disregarding groups of $N_c$ elements. In more mathematical terms, one can define $\nu_R$ as the remainder in the Euclidean division by $N_c$ of the number of elements of ${\cal T}_R$. It is thus an integer comprised between $0$ and $N_c-1$. 

It is convenient in practice to lift this restriction by stating that two integers that differ by a multiple of $N_c$ represent the same $N_c$-ality, that is to define $\nu_R$ as the number of elements of ${\cal T}_R$, modulo $N_c$. In mathematical terms,\footnote{Strictly speaking, one should denote the elements of $\mathds{Z}/N_c\mathds{Z}$ using a different notation, for instance $\dot{\nu}_R$, to indicate that this represents a class of elements of $\mathds{Z}$, the class of elements of the form $\nu_R+N_ck$, with $\smash{k\in\mathds{Z}}$. We will avoid this extra layer of notation, however. When there could be any kind of ambiguity, we will resort to the notation ${\rm mod}\,N_c$ to indicate that a given quantity/equation is defined/applies only modulo $N_c$.} $\nu_R$ is an element of $\mathds{Z}/N_c\mathds{Z}$. This point of view, which we shall adopt in what follows, allows one to add $N_c$-alities together very naturally. We will also compare $N_c$-alities with each other, but this will be done by first bringing them between $0$ and $N_c-1$.

As a trivial example, the $N_c$-ality of the fundamental representation $\nu$ is nothing but $\nu$, that is $\smash{\nu_\nu=\nu}$. Reciprocally, and unless otherwise stated, the label $\nu$ used for fundamental representations will also be promoted into an element of $\mathds{Z}/N_c\mathds{Z}$ in such a way that $\nu$ and $\nu+N_c k$ represent one and the same fundamental representation. Note that there is no fundamental representation corresponding to $\smash{\nu=0}$. However, it is quite natural and convenient to use $\smash{\nu=0}$ to denote the trivial or singlet representation ${\bf 1}$. The latter is an irreducible representation of dimension $1$ which associates the identity transformation to each element of the group.

As for reducible representations, if $R$ is the tensor product of two irreducible representations $R_1$ and $R_2$, then its $N_c$-ality is defined as the sum (within $\mathds{Z}/N_c\mathds{Z}$) of the $N_c$-alities of each of the factors:
\beq
\nu_{R_1\otimes R_2}\equiv\nu_{R_1}+\nu_{R_2}\,.
\eeq
If $R$ is instead the direct sum of $R_1$ and $R_2$, there is no natural way to associate an $N_c$-ality to $R$, except when the $N_c$-alities of $R_1$ and $R_2$ are equal. In this case, one sets
\beq
\nu_{R_1\oplus R_2}\equiv \nu_{R_1}=\nu_{R_2}\,.
\eeq
We will see one example below.

One reason for classifying the various SU($N_c$) representations according to their $N_c$-ality is that, in the absence of quarks, all Polyakov loops associated to representations with equal $N_c$-ality transform in the same way under center transformations\footnote{In the presence of quarks, in addition to this phase multiplication, the boundary conditions for the quarks in the Euclidean functional integral formulation (\ref{eq:lR}) are changed from antiperiodic to antiperiodic modulo $e^{i2\pi/N_c}$, or, equivalently, the chemical potential is shifted by $i(2\pi/N_c)T$.}
\beq
\ell_R\to e^{-i2\pi \nu_R/N_c}\ell_R\,.\label{eq:cs}
\eeq
We refer to App.~\ref{app:pot} for a proof of this fact.

\subsection{Low-temperature behavior}
Let us now argue that the low-temperature behavior of the non-fundamental Polyakov loops is directly connected to the $N_c$-ality of the corresponding representations. It will be important to keep in mind that, unlike the fundamental Polyakov loops, the non-fundamental ones do not derive directly from a potential but are, instead, expressed in terms of the fundamental Polyakov loops, which we represent by the functional notation $\ell_R(\ell_\nu)$.

At low temperatures and for $\smash{|\mu|<M}$, since the $\ell_\nu$ are exponentially suppressed, we can consider a Taylor expansion
\beq
\ell_R(\ell_\nu)=\sum_{p\geq 0}\frac{1}{p!}\frac{\partial^{p+1}\ell_R}{\partial\ell_{\nu_1}\cdots\ell_{\nu_p}}\ell_{\nu_1}\cdots\ell_{\nu_p}\,,
\eeq
where it is implicitly understood that the derivatives are taken at $\smash{\ell_\nu=0}$. Also, because we have assumed that quarks decouple at low temperatures, the coefficients $\partial^{p+1}\ell_R/\partial\ell_{\nu_1}\cdots\ell_{\nu_p}$ become those of the purely gluonic theory, in which case center symmetry (\ref{eq:cs}) implies
\beq
\ell_R(e^{-i2\pi\nu/N_c}\ell_\nu)=e^{-i2\pi\nu_R/N_c}\ell_R(\ell_\nu)\,.\label{eq:54}
\eeq
It follows from this that
\beq
& & e^{-i2\pi(\nu_1+\cdots+\nu_p-\nu_R)/N_c}\frac{\partial^{p+1}\ell_R}{\partial\ell_{\nu_1}\cdots\ell_{\nu_p}}=0\,,
\eeq
and thus that any derivative that does not satisfy
\beq
\nu_1+\cdots+\nu_p=\nu_R\mod{N_c}\,,
\eeq
needs to vanish. In other words, the products of fundamental Polyakov loops that contribute to the Taylor expansion of $\ell_R(\ell_\nu)$ are necessarily such that their $N_c$-alities add up to the $N_c$-ality of $R$. 

If we now recall that any product $\ell_{\nu_1}\cdots\ell_{\nu_p}$ with $\smash{p\geq 2}$ is suppressed with respect to $\ell_{\nu_R}$, see Eqs.~(\ref{eq:29}) and (\ref{eq:30}), and because there is a priori no reason (no symmetry) for the linear term to be absent, the Taylor expansion is dominated by the linear approximation:
\beq
\ell_R(\ell_\nu)-\ell_R(0)\simeq \frac{\partial\ell_R}{\partial\ell_{\nu_R}}\ell_{\nu_R}\,.
\eeq
If $\smash{\nu_R\neq 0}$, then $\smash{\ell_R(0)=0}$ from Eq.~(\ref{eq:54}), and thus
\beq
\ell_R\simeq \frac{\partial\ell_R}{\partial\ell_{\nu_R}}\ell_{\nu_R}\,,\label{eq:propto}
\eeq
with a pre-factor $\partial\ell_R/\partial\ell_{\nu_R}$ that does not depend on $\mu$ at low temperatures. This implies
\beq
\Delta Q_R\sim \Delta Q_{\nu_R}\,.\label{eq:conjecture2}
\eeq
Similarly, if $\smash{\nu_R=0}$, then $\smash{\ell_R\to\ell_R(0)\neq 0}$, with $\ell_R(0)$ independent of $\mu$ at low temperatures. It follows that $\smash{\Delta Q_R\to 0}$, which is again compatible with (\ref{eq:conjecture2}), since $\smash{\Delta Q_0=0}$.

We shall then retain that, if no particular symmetry or additional constraint makes the derivative $\partial\ell_R/\partial\ell_{\nu_R}$ vanish at the confining point (and if this derivative is dominated by pure gauge contributions at low temperatures), then the net quark number gained by the bath upon bringing a probe in an irreducible color representation $R$ is the same as the net quark number gained by the bath when the probe is in the fundamental representation $\nu_R$, that is in the fundamental representation of same $N_c$-ality. As before, we do not know the particular color representation of the states that are formed to absorb the probe, but we know that they necessarily have a vanishing $N_c$-ality, because $\nu_R+\Delta Q_R=\nu_R+\Delta Q_{\nu_R}=0$ or $N_c$, see Secs.~\ref{sec:color} for further discussions in relation with color.

As for the free-energy gains, the question is more delicate as it depends on the behavior of $\partial\ell_R/\partial\ell_{\nu_R}$ at low temperatures. If the assumed glue dominance in $\partial\ell_R/\partial\nu_R$ stems from the presence of power-law contributions (as in most models for the glue sector), the contribution of this prefactor to the free-energy at low temperatures is negligible, and we find
\beq
\Delta F_R\sim\Delta F_{\nu_R}\,.\label{eq:conjecture3}
\eeq
If, instead, $\partial\ell_R/\partial\ell_{\nu_R}$ is exponentially suppressed then
\beq
\Delta F_R>\Delta F_{\nu_R}\,.\label{eq:conjecture4}
\eeq
We stress that numerical simulations of heavy-quark QCD at finite chemical potential can be used to evaluate $\ell_R$, $\Delta F_R$ and $\Delta Q_R$, and to test predictions such as (\ref{eq:conjecture2}), (\ref{eq:conjecture3}) or (\ref{eq:conjecture4}), and thus to test various of the considered assumptions, in the glue sector in particular.

\section{Tree-level Polyakov loops}\label{sec:tree}

 Within appropriately chosen gauges, the behavior of the Polyakov loops is well captured in terms of the gauge-field expectation value in that gauge, which serves as a semiclassical background. At leading, tree-level order of a semi-classical expansion around that background, the relations between the fundamental and non-fundamental Polyakov loops are then easily obtained and mimic the relations obeyed by the characters of SU($N_c$), see App.~\ref{app:character} for more details. This allows one to determine the tree-level non-fundamental Polyakov loops in terms of the fundamental ones. Let us now investigate whether the so determined tree-level Polyakov loops comply with the above general expectations.

\subsection{Character identities}
Consider two irreducible representations $R_1$ and $R_2$ of SU($N_c$) whose tensor product decomposes as a direct sum of other irreducible representations $R'_i$:
\beq
R_1\otimes R_2={\bigoplus}_i R'_i\,.
\eeq
The corresponding tree-level Polyakov loops then satisfy the character identity
\beq
d_{R_1}d_{R_2}\ell_{R_1}\ell_{R_2}={\sum}_i d_{R'_i}\ell_{R'_i}\,,\label{eq:char}
\eeq
see Ref.~\cite{Gupta:2007ax} and App.~\ref{app:character}. Note that the various dimensions involved are constrained by
\beq
d_{R_1}d_{R_2}={\sum}_i d_{R'_i}\,.
\eeq  
Moreover, the $N_c$-ality of each irreducible representation appearing in the decomposition is fixed by the $N_c$-ality of the original tensor product:\footnote{This is then a case where one can give a meaning to $\nu_{\bigoplus_i R'_i}$, see the discussion above.}
\beq
\forall i,\,\,\nu_{R'_i}=\nu_{R_1\otimes R_2}\equiv\nu_{R_1}+\nu_{R_2}\,.
\eeq
This last property is rooted in the very way decompositions such as (\ref{eq:char}) are implemented using the notion of Young tableaux: the total number of elements of the Young tableaux of $R_1$ and $R_2$ is the total number of elements that will enter the Young tableaux associated to each of the $R'_i$, modulo $N_c$, see Ref.~\cite{Gernot} for a pedagogical exposure on the use of Young tableaux.

For formulas such as (\ref{eq:char}) to give access to all non-fundamental Polyakov loops in terms of the fundamental ones, one should use them in a specific order such that, at each step, only one new non-fundamental representation is involved. Here, we shall consider one such possible order that fits our needs. We shall also use $(\nu_1;\nu_2;\cdots;\nu_p)$, with $\smash{\nu_1\geq\nu_2\geq\cdots\geq\nu_p}$ to denote the irreducible representation associated with the $p$-column Young tableau with  $\nu_i$ elements in the $i$-th column. This is certainly not the most compact notation to represent irreducible representations, but again, a convenient one for our purpose. The $\nu_i$ will be allowed to exceed $N_c$, but if one of them does so strictly, there is actually no Young tableau and no contribution to the tensor decomposition. When all the $\nu_i$ are smaller than $N_c$ but some of them equal $N_c$, the Young tableau has as many fewer columns. When all the columns need to be removed, one obtains the trivial, identity representation, which we denote by $\smash{\nu=0}$ or $\smash{\nu=N_c}$. The associated tree-level Polyakov loop is $\smash{\ell_{N_c}=\ell_0=1}$.

\begin{figure}[t]
    \centering
    \includegraphics[width=0.9\linewidth]{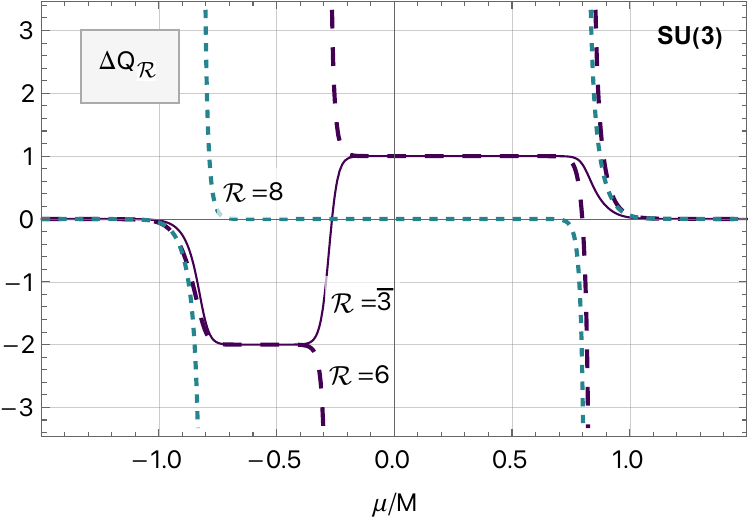}
    \caption{Net quark number gains associated with the SU(3) representations of type $(\nu;1)$ (medium/long dashed) compared to those of the corresponding fundamental representations of equal $N_c$-ality $\nu+1$ (thin).}
    \label{fig: su3 nonfund irred DQ}
\end{figure}

\subsection{Young tableaux of type $(\nu;1)$}

We start by considering tensor products between the fundamental representations $\smash{\nu\geq 1}$ and $\smash{\nu=1}$. They decompose as
\beq
\nu\otimes 1=(\nu+1)\oplus (\nu;1)\,,\label{eq:nu1}
\eeq
where $\nu+1$ is a fundamental representation, or the trivial representation if $\smash{\nu=N_c-1}$. Thus, Eq.~(\ref{eq:nu1}) allows one to construct all the non-fundamental representations $(\nu;1)$ from the knowledge of the fundamental ones and the trivial one.

Let us now translate these considerations in terms of non-fundamental Polyakov loops at tree-level. From Eqs.~(\ref{eq:char}) and (\ref{eq:nu1}), we deduce that
\beq
d_\nu d_1\ell_\nu\ell_1=d_{\nu+1}\ell_{\nu+1}+d_{(\nu;1)}\ell_{(\nu;1)}\,.
\eeq
Using Eqs.~(\ref{eq:29}) and (\ref{eq:30}), we can neglect $\ell_\nu\ell_1$ against $\ell_{\nu+1}$ and thus
\beq
\ell_{(\nu;1)}\sim-\frac{d_{\nu+1}}{d_{(\nu;1)}}\ell_{\nu+1}\,,\label{eq:1st}
\eeq
from which it follows that
\beq
\Delta Q_{(\nu;1)}\sim\Delta Q_{\nu+1}\,.
\eeq
We have thus found that the net quark number gain associated to the non-fundamental representation $(\nu;1)$ is the same as that of the fundamental representation of same $N_c$-ality, in line with the general expectation (\ref{eq:conjecture2}).

\begin{figure}[t]
    \centering
    \includegraphics[width=0.9\linewidth]{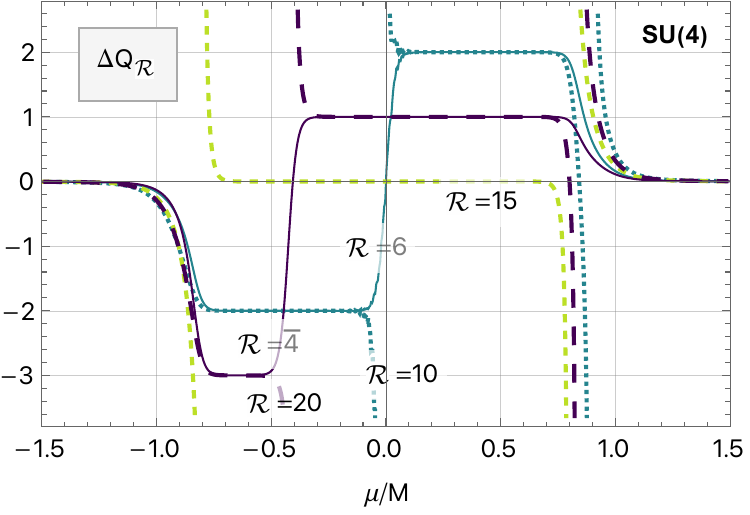}
    \caption{Net quark number gains associated with the SU(4) representations of type $(\nu;1)$ (short/medium/long dashed) compared to those of the corresponding fundamental representations of equal $N_c$-ality $\nu+1$ (thin).}
    \label{fig: su4 nonfund irred DQ}
\end{figure}

In the SU(3) case, for instance, there are two representations of the type $(\nu;1)$. These are ${\bf 6}$ and ${\bf 8}$, as generated from the tensor products
\begin{equation}
    {\bf 3}\otimes {\bf 3}={\bf \bar{3}}\oplus {\bf 6} \quad {\rm and \quad }{\bf \bar{3}}\otimes {\bf 3}={\bf 1}\oplus {\bf 8}\,.
\end{equation}
The comparison of their net quark number gains with $\Delta Q_2$ (${\bf\bar{3}}$)  and $\Delta Q_3=\Delta Q_0=0$ (${\bf 1}$) respectively is shown in Fig.~\ref{fig: su3 nonfund irred DQ} in the case of the model of Ref.~\cite{MariavanEgmond2022ATemperature}. Similarly, in the SU(4) case, there are three such representations, ${\bf 10}$, ${\bf 20}$ and ${\bf 15}$, as generated from the tensor products
\begin{equation}
    {\bf 4}\otimes {\bf 4}={\bf 6}\oplus {\bf 10}\,,\,\,\, {\bf 6}\otimes {\bf 4}={\bf \bar{4}}\oplus {\bf 20}\,,\,\,\, {\bf \bar{4}}\otimes {\bf 4}={\bf 1}\oplus {\bf 15}\,.
\end{equation}
The comparison of their net quark number gains with $\Delta Q_2$ (${\bf 6}$), $\Delta Q_3$ (${\bf\bar{4}}$) and $\Delta Q_4=\Delta Q_0=0$ (${\bf 1}$) respectively is shown in Fig.~\ref{fig: su4 nonfund irred DQ}.

Note that, in both Fig.~\ref{fig: su3 nonfund irred DQ} and Fig.~\ref{fig: su4 nonfund irred DQ}, in addition to the expected plateaux for $\Delta Q_{(\mu;1)}$, we find divergences located at certain values of $\mu$. We believe that these are artifacts of the tree-level approximation. In fact, from Eq.~(\ref{eq:1st}), we see that, the latter leads to a negative Polyakov loop $\ell_{(\nu;1)}$, which is obviously in tension with the interpretation (\ref{eq:FR}) in terms of a (real) free-energy. Even though these imaginary contributions to the free-energy do not affect the extraction of the net quark number gains, and vanish at low temperatures,\footnote{The reason why we still see their effect in Fig.~\ref{fig: su3 nonfund irred DQ} and Fig.~\ref{fig: su4 nonfund irred DQ} is that, numerically, it is quite hard to take the limit $\smash{T\to 0}$ strictly.} they need to be seen as a limitation of the tree-level approximation. We will see one more limitation in the next section.

\subsection{Young tableaux of type $(\nu;2)$}
Next one considers tensor products between $\smash{\nu\geq 2}$ and $\smash{\nu=2}$. They decompose as
\beq
\nu\otimes 2=(\nu+2)\oplus(\nu+1;1)\oplus (\nu;2)\,.\label{eq:nu2}
\eeq
For any value of $\smash{2\leq\nu\leq N_c-1}$, the last representation appearing in the decomposition is a new one, in the sense that it has two columns, with two elements in the last column. The other two representations are already known from the previous step. More precisely, for $\smash{\nu\leq N_c-3}$ (this requires $\smash{N_c\geq 5}$ since $\smash{\nu\geq 2}$), the first one is a fundamental representation and the second a two-column representation with one element in the second column, for $\smash{\nu=N_c-2}$ (this requires $\smash{N_c\geq 4}$), the first one becomes the trivial representation, while for $\smash{\nu=N_c-1}$ (this requires $\smash{N_c\geq 3}$), the first one is absent while the second becomes the fundamental representation $\smash{\nu=1}$.

Despite these various cases, it is always true that
\beq
\nu\otimes 2=((\nu+1)\otimes 1)\oplus (\nu;2)\,.\label{eq:eq3}
\eeq
Using the character identities, we deduce that
\beq
\ell_{(\nu;2)}\sim \frac{d_\nu d_2}{d_{(\nu;2)}} \ell_\nu\ell_2-\frac{d_{\nu+1}d_1}{d_{(\nu;2)}}\ell_{\nu+1}\ell_1\,.\label{eq:temp}
\eeq
In the case $\smash{\nu=N_c-1}$, this becomes
\beq
\ell_{(N_c-1;2)}\sim \frac{d_{N_c-1} d_2}{d_{(N_c-1;2)}} \ell_{N_c-1}\ell_2-\frac{d_1}{d_{(N_c-1;2)}}\ell_1\,.
\eeq
Since $\ell_{N_c-1}\ell_2$ is suppressed with respect to $\ell_1$, we deduce once more that
\beq
\ell_{(N_c-1;2)}\sim-\frac{d_1}{d_{(N_c-1;2)}}\ell_1\,,
\eeq
and then
\beq
\Delta Q_{(N_c-1;2)}\sim\Delta Q_1\,,
\eeq
in line once more with (\ref{eq:conjecture2}).

In the case $\smash{\nu\leq N_c-2}$, using similar arguments as those in App.~\ref{app:ordering}, it is easily argued that the first term in the right-hand side of Eq.~(\ref{eq:temp}) is suppressed or is of the same order than the second one at low temperatures and $\smash{|\mu|<M}$. Then
\beq
\ell_{(\nu;2)}\propto\ell_{\nu+1}\ell_1\,,
\eeq
with a prefactor that does not depend on $\mu$ at low temperatures. It follows this time that
\beq
\Delta Q_{(\nu;2)}=\Delta Q_{\nu+1}+\Delta Q_1\,,\label{eq:ko1}
\eeq
which violates the general expectation (\ref{eq:conjecture2}). As we will explain in Sec.~\ref{sec:beyond}, this violation could originate in the considered tree-level approximation, which introduces additional relations between the Polyakov loops.

\begin{figure}[t]
    \centering
    \includegraphics[width=0.9\linewidth]{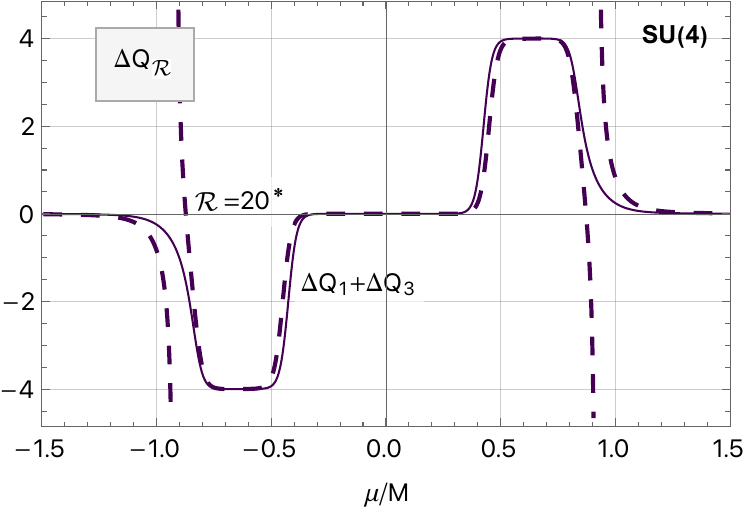} 
    \caption{Net quark number gains associated to the SU(4) representation ${\bf 20^*}$ (dashed) and comparison with $\Delta Q_1+\Delta Q_3$ (thin).}
    \label{fig:ko}
\end{figure}

In the SU(3) case, there is only one representation of the type $(\nu;2)$, ${\bf \bar{6}}$, and because it is the charge-conjugate of ${\bf 6}$, a representation of type $(\nu;1)$, it makes no doubt that it obeys (\ref{eq:conjecture2}). In the SU(4) case, in contrast, there are two representations of the type $(\nu;2)$, ${\bf \overline{20}}$ and ${\bf 20^*}$. As the charge conjugate of ${\bf 20}$, the first one fulfills (\ref{eq:conjecture2}). However, ${\bf 20^*}$, which originates from the product
\begin{equation}
    {\bf 6}\otimes {\bf 6}={\bf 1}\oplus{\bf 15}\oplus {\bf 20^*}
\end{equation}
fulfills (\ref{eq:ko1}) instead, as we illustrate in Fig.~\ref{fig:ko}. We stress again that we view this as a peculiarity of the tree-level approximation. We note again the presence of divergences in $\Delta Q_{(\nu;2)}$ for certain values of $\mu$ if the temperature is not taken strictly to $0$. These originate in the Polyakov loops $\ell_{(\nu;2)}$ becoming negative at low temperatures, a limitation of the tree-level approximation.\footnote{It could also very well be that this is an indication that the non-fundamental Polyakov loops do not admit the free-energy interpretation (\ref{eq:FR}), see also the discussion in footnote 8.}

In what follows, rather than pursuing the investigation of the tree-level Polyakov loops associated with even higher Young tableaux, we discuss how these limitations could potentially be cured by higher-order corrections beyond tree-level.

\subsection{Beyond tree-level}\label{sec:beyond}
Beyond tree-level, it is not true anymore that $\ell_{R_1\otimes R_2}=\ell_{R_1}\ell_{R_2}$. This means that the best that one can deduce from the character identities is that
\beq
d_{R_1}d_{R_2}\ell_{R_1\otimes R_2} = \sum_i d_{R'_i}\ell_{R'_i}\,.\label{eq:char2}
\eeq
Concomitantly, there is no reason not to consider the presence of a linear contribution $\ell_{\nu_1+\nu_2}$ in the low temperature behavior of $\ell_{R_1\otimes R_2}$: 
\beq
\ell_{R_1\otimes R_2}-\ell_{R_1\otimes R_2}(0)\sim\frac{\partial\ell_{R_1\otimes R_2}}{\partial\ell_{\nu_1+\nu_2}}\ell_{\nu_1+\nu_2}\,.\label{eq:eq}
\eeq
Let us now see how this can propagate to the Polyakov loops in all representations, to ensure that (\ref{eq:conjecture2}) is fulfilled while potentially solving the sign problem referred to in the previous section. For convenience, we will summarize the relevant information contained in (\ref{eq:eq}) as
\beq
\ell_{R_1\otimes R_2}\sim\frac{\partial\ell_{R_1\otimes R_2}}{\partial\ell_{\nu_1+\nu_2}}\ell_{\nu_1+\nu_2}\,,\label{eq:eq2}
\eeq
where we have set $\smash{\partial X/\partial\ell_0\equiv X(0)}$.

Let us first consider the representations of the type $(\nu;1)$. From Eqs.~(\ref{eq:nu1}) and (\ref{eq:char2}), we have
\beq
d_\nu d_1\ell_{\nu\otimes 1}=d_{\nu+1}\ell_{\nu+1}+d_{(\nu;1)}\ell_{(\nu;1)}\,.
\eeq
Using Eq.~(\ref{eq:eq2}), we obtain
\beq
\ell_{(\nu;1)}\sim\frac{d_\nu d_1\frac{\partial\ell_{\nu\otimes 1}}{\partial\ell_{\nu+1}}-d_{\nu+1}}{d_{(\nu;1)}}\ell_{\nu+1}\,.\label{eq: 1st}
\eeq
Unless the prefactor vanishes, it follows that Eq.~(\ref{eq:conjecture2}) is verified for any representation $(\nu;1)$, just as we saw before, but this time there is room for $\ell_{(\nu;1)}$ to be positive, unlike what happened earlier.

Consider next the representations of the type $(\nu;2)$. From Eq.~(\ref{eq:eq3}), we have
\beq
d_\nu d_2\ell_{\nu\otimes 2}=d_{\nu+1}d_1\ell_{(\nu+1)\otimes 1}+d_{(\nu;2)}\ell_{(\nu;2)}\,,\nonumber\\
\eeq
and thus
\beq
\ell_{(\nu;2)}\sim \frac{d_\nu d_2\frac{\partial\ell_{\nu\otimes 2}}{\partial\ell_{\nu+2}}-d_{\nu+1}d_1\frac{\partial\ell_{(\nu+1)\otimes 1}}{\partial\ell_{\nu+2}}}{d_{(\nu;2)}}\ell_{\nu+2}\,,
\eeq
again compatible with (\ref{eq:conjecture2}) unless the prefactor vanishes.

By iterating this procedure, one finds
\beq
\nu\otimes\nu'=((\nu+1)\otimes(\nu'-1))\oplus (\nu;\nu')\,,\label{eq:nunu'}
\eeq
for $\smash{\nu\geq\nu'}$. It follows that
\beq
d_\nu d_{\nu'}\ell_{\nu\otimes \nu'}=d_{\nu+1}d_{\nu'-1}\ell_{(\nu+1)\otimes \nu'-1}+d_{(\nu;\nu')}\ell_{(\nu;\nu')}\,,\nonumber\\
\eeq
and thus
\beq
\ell_{(\nu;\nu')}\sim \frac{d_\nu d_{\nu'}\frac{\partial\ell_{\nu\otimes \nu'}}{\partial\ell_{\nu+\nu'}}-d_{\nu+1}d_{\nu'-1}\frac{\partial\ell_{(\nu+1)\otimes (\nu'-1)}}{\partial\ell_{\nu+\nu'}}}{d_{(\nu;\nu')}}\ell_{\nu+\nu'}\,,\nonumber\\
\eeq
which implies (\ref{eq:conjecture2}) unless the pre-factor vanishes.

One could proceed in a similar way to analyze higher column Young tableaux. In the next step, we consider all tensor products $(\nu;\nu')\otimes 1$ with $\smash{\nu\geq\nu'}$. Their decompositions involve a number of representations associated with already generated, one- or two-column Young tableaux, plus a single representation associated with the three-column Young tableau $(\nu;\nu';1)$. Next one considers all tensor products $(\nu;\nu')\otimes 2$ with $\smash{\nu\geq\nu'\geq 2}$. Their decompositions involve some representations associated with already generated, one- or two-column Young tableaux, or three-column Young tableaux with one element in the third column, plus a single, new representation associated with the three-column Young tableau $(\nu;\nu';2)$. By iterating this procedure, one obtains $(\nu;\nu';\nu'')$ as the result of removing from $(\nu;\nu')\otimes\nu''$ all representations $\rho$, $(\rho;\rho')$ and $(\rho;\rho';\rho'')$ with $\smash{\rho''<\nu''}$. Further iterating this construction, the irreducible representation $(\nu_1;\cdots;\nu_p)$ can be obtained by removing from $(\nu_1;\cdots;\nu_{p-1})\otimes\nu_p$ all the representations associated with $(p-1)$-column Young tableaux or with $p$-column Young tableaux with less than $\nu_p$ elements in the last column. Since all the representations that appear in a given step have the same $N_c$-ality, and thus their Polyakov loops scale as the associated fundamental one with a $\mu$-independent pre-factor, so does the Polyakov loop of the newly generated representation (if no weird cancellations occur), and the corresponding net quark number gain is that of the associated fundamental representation.

\section{Color}\label{sec:color}
Let us return to the question of color. As mentioned above, we cannot determine the color representation of the states that are formed to absorb the probe, and, in particular, whether those are color-singlet states. We can at best say that the formed states have a vanishing $N_c$-ality. Another thing we can do is to determine in which cases color-singlet states involving a minimal number of quarks and/or antiquarks are ruled out. This relates to the way Young tableaux are combined to put representations together.

Take for instance the case $\smash{\mu>\mu_{\nu_R}}$. According to (\ref{eq:conjecture2}), the system brings $N_c-\nu_R$ quarks, but in order for these quarks to combine with the probe in representation $R$ to form a color-singlet state, we need these quarks to complete all columns of the corresponding Young tableau into full columns of size $N_c$. If we denote by $\nu_i$ the number of elements of column $i$, this means that we should have
\beq
\sum_{i=1}^p(N_c-\nu_i)= N_c-\nu_R\,.\label{eq:condition}
\eeq
To continue, it is convenient to introduce the complement of $\nu$ as $\nu'\equiv N_c-\nu$. The above condition rewrites
\beq\label{eq: complement cond}
\sum_{i=1}^p \nu'_i=\nu'_R\,,
\eeq
and expresses a constraint between the complement of the $N_c$-ality of $R$, and the complement of the sizes of the columns of its Young tableau. The left-hand side is bounded from below by $p$, which means that a necessary condition for (\ref{eq:condition}) to hold true is $\smash{p\leq \nu'_R}$. Similarly, the left-hand side of (\ref{eq:condition}) is bounded from below by $p-1+\nu'_i$ for any $i$ which means that a necessary condition for (\ref{eq:condition}) to hold true is $\nu'_i\leq\nu'_R-p+1$. Thus, among all probes of a given complement $N_c$-ality $\nu'$, those which can form color-singlet states with the $N_c-\nu'$ quarks of the bath need to have less than $\nu'$ columns, each with a complement column of size smaller than $\nu'-p+1$. 

\begin{figure}
    \centering
    \includegraphics[width=0.75\linewidth]{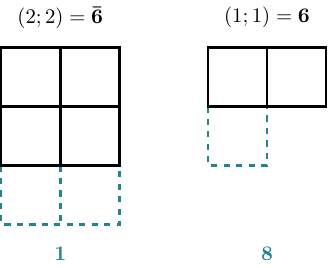}
    \caption{Examples of SU(3) Young tableaux for two representations with $\smash{\nu_{\mathbf{\bar{6}}}=1}$ (left) and $\smash{\nu_{\mathbf{6}}=2}$ (right), with the dashed lines showing possible distributions of the quark responses, i.e., the complements $\smash{\nu_{\mathbf{\bar{6}}}^\prime=2}$, $\smash{\nu_{\mathbf{6}}^\prime=1}$ to the 3-ality. $\mathbf{\bar{6}}$ fulfills Eq.~\eqref{eq: complement cond}, but $\mathbf{6}$ does not, so to form a color-singlet state the medium must provide another state of vanishing 3-ality, like a gluon, which the minimal number of quarks compatible with the net quark number gain cannot account for.}
    \label{fig: color neutrality}
\end{figure}

There is only one such Young tableau with $\nu'$ columns, each of size $N_c-1$. This tableau automatically satisfies (\ref{eq:condition}). The left diagram of Fig.~\ref{fig: color neutrality} shows an example of such a tableau for SU(3). If we decrease the number of columns by one unit, we should increase one of the $\nu'_i$ by one unit to still fulfill (\ref{eq:condition}).  If we further decrease the number of columns by one unit, we can increase that same $\nu'_i$ by one extra unit, or a new one by one unit, and so on until we reach a Young tableau with only one column, in which case the corresponding $\nu'_1$ has to be equal to $\nu'$.

Any Young tableau that does not fit into this classification cannot form a color-singlet representation with the $N_c-\nu_R$ quarks from the bath. Still, this does not necessarily mean that the probe is screened into a non-singlet representation. The point is that this state has a vanishing $N_c$-ality and could thus combine with gluons, which also have a vanishing $N_c$-ality, or equivalently, with additional quark/antiquark pairs (which do not change $\Delta Q_R$) in the adjoint representation. The adjoint representation corresponds to the Young tableau  $(N_c-1;1)$, so the scenario in which the probe in representation $R$ is absorbed into a color-singlet state involving $N_c-\nu_R$ quarks from the bath, as well as some gluons from the bath, requires that the tensor product
\beq
{\otimes}^k (N_c-1;1)\otimes R\,{\otimes}^{N_c-\nu_R} 1,
\eeq
contains the trivial representation in its decomposition, for some $k$. We have used the notation ${\otimes}^kR$ to represent the tensor product of $k$ representations $R$. The discussion of whether this is always the case for representations $R$ that do not fulfill (\ref{eq:condition}) is left for a future study. Instead, we consider an illustrative example in the SU(3) case, sketched on the right of Fig.~\ref{fig: color neutrality}. At large chemical potential, a probe made of two quarks in representation ${\bf 6}$ is screened by one quark of the medium, see also Fig.~\ref{fig: su3 nonfund irred DQ}. Since 
\beq
{\bf 6}\otimes {\bf 3}={\bf 8}\oplus{\bf 10}\,,\label{eq:dec}
\eeq
this system of three quarks cannot be in a color-singlet state. This happens because the Young tableau of ${\bf 6}$, a one-line tableau with two elements, does not fulfill the necessary condition $\smash{\nu'_i\leq\nu_R-p+1}$. Indeed, $\smash{\nu'_i=2}$ whereas $\smash{\nu_R-p+1=1}$ in this case. Still, we note that the decomposition (\ref{eq:dec}) contains ${\bf 8}$, which means that the formed state can still form a color-singlet state upon combination with a gluon or a quark/antiquark pair in the adjoint representation. Indeed,
\beq
{\bf 8}\otimes {\bf 8}={\bf 1}\oplus{\bf 8}\oplus{\bf 8}\oplus{\bf 10}\oplus{\bf \overline{10}}\oplus{\bf 27}\,,
\eeq
contains the singlet representation ${\bf 1}$ in its decomposition.

\section{Minkowskian formulation}\label{sec:heuristic}

The results we have derived so far are based on the Polyakov loop and on the associated center symmetry. These are intrinsically Euclidean notions as they appear in the Euclidean functional integral formulation of QCD. It is then interesting to wonder whether one can recover the above results from a Minkowskian perspective, while investigating the status of the Polyakov loop and of center symmetry within this framework.

In a Minkowskian formulation, the Polyakov loop (\ref{eq:lR}) is meant to correspond to the ratio of two partition functions,
\beq
\ell_R=\frac{Z_R}{Z}\,,\label{eq:lH}
\eeq
the partition function
\beq
Z_R\equiv {\rm Tr}_R\,e^{-\beta(H-\mu Q)}\label{eq:ZR}
\eeq
of a thermal bath of quarks and gluons in the presence of a heavy color probe in representation $R$, and the partition function
\beq
Z\equiv{\rm Tr}\,e^{-\beta(H-\mu Q)}\label{eq:Z}
\eeq
of that same bath in the absence of a probe. The partition functions explicitly involve the Hamiltonian of the system, which is derived from the Minkowskian QCD action.

\subsection{Regarding center symmetry}
One difficulty with the above picture, however, is that it is far from obvious how the center symmetry and the associated transformation of $\ell_R$ are encoded in the Minkowskian formulation (\ref{eq:lH})-(\ref{eq:Z}). In fact, center symmetry is usually defined within the Euclidean functional integral formulation of QCD where the traces in Eqs.~(\ref{eq:ZR}) and (\ref{eq:Z}) are replaced by functional integrals involving an Euclidean version of the gauge field, which is itself a function of an Euclidean, compact time of extent $\smash{\beta\equiv 1/T}$. These two notions, which enter the definition of center transformations, are absent from the Minkowskian formulation and there is then no obvious way to define how a center transformation acts on the states of the Hilbert space.

Another puzzling fact is that center transformations typically multiply the Polyakov loop by a center element $e^{-i2\pi/N_c\nu_R}$, see App.~\ref{app:pot}, that is by a complex number, while the Minkowskian formulation (\ref{eq:lH})-(\ref{eq:Z}) in terms of traces of hermitian operators leaves no room for complex numbers. One should stress, however, that, within the Euclidean functional integral framework, and concomitantly with the multiplication by the factor $e^{-i2\pi/N_c\nu_R}$, center transformations also modify the boundary conditions of the quarks, from antiperiodic ones to antiperiodic modulo a center element \cite{MariSurkau2025Heavy-quarkModel}. Since the traces of the Minkowskian formulation are related to the field boundary conditions in the Euclidean functional integral formulation, the change of quark boundary conditions upon a center transformation in the Euclidean formulation implies that a proper account of center symmetry within the Minkowskian formulation requires one to embed the latter into a larger framework where, in particular, the traces in Eqs.~(\ref{eq:ZR})-(\ref{eq:Z}) would be replaced by a more general notion.

Finding this more general framework lies beyond the scope of the present work, but we can proceed in a different, yet instructive way. To this purpose, we recall that, within the Euclidean functional integral framework and upon combining a center transformation with complex conjugation, see e.g. \cite{MariSurkau2025Heavy-quarkModel}, it is possible to restore the antiperiodic boundary conditions of the quarks, at the price of shifting the quark chemical potential by $i(2\pi/N_c)T$.\footnote{This shift is also the one associated to the Roberge-Weiss transition, see, however, the discussion in Ref.~\cite{MariSurkau2025Heavy-quarkModel}.} An immediate consequence of this is that the Polyakov loops should obey the identity
\beq
\ell_R\left(\mu+i\frac{2\pi}{N_c}T\right)=e^{-i\frac{2\pi}{N_c}\nu_R}\ell_R(\mu)\,.\label{eq:lRt}
\eeq
Thus, if the ratio of traces in Eq.~(\ref{eq:lH}) is to faithfully represent the Polyakov loop, it should also obey the identity (\ref{eq:lRt}).

\subsection{State constraints}
Because $H$ and $Q$ commute with each other, we can write
\beq
Z_R={\rm Tr}_R\,e^{-\beta H}e^{\beta\mu Q}\,,
\eeq
and
\beq
Z={\rm Tr}\,e^{-\beta H}e^{\beta\mu Q}\,.
\eeq
Imagine now evaluating the traces ${\rm Tr}_R$ and ${\rm Tr}$ by summing over a basis of states $|\psi\rangle$ that diagonalize $Q$ in the subspace associated to the presence or absence of the probe respectively. If we assume that all these states satisfy
\beq
Q|\psi\rangle=(-\nu_R+N_c k)|\psi\rangle\,,\label{eq:conj}
\eeq
and\footnote{Note that the condition for the states that enter the trace ${\rm Tr}$ corresponding to the absence of probe is the same than the condition for the states that enter the trace ${\rm Tr}_R$ corresponding to the presence of a probe in a representation with vanishing $N_c$-ality. This does not mean, however, that the traces are the same.}
\beq
Q|\psi\rangle=N_c k|\psi\rangle\,,\label{eq:conj2}
\eeq
respectively, for some integer $k$, then it is easily checked that
\beq
Z_R\left(\mu+i\frac{2\pi}{N_c}T\right)=e^{-i\frac{2\pi}{N_c}\nu_R}Z_R(\mu)\,,
\eeq
and
\beq
Z(\mu+i2\pi/N_cT)=Z(\mu)\,, 
\eeq
from which (\ref{eq:lRt}) follows immediately. 

Equation (\ref{eq:conj}), including (\ref{eq:conj2}) for $\smash{\nu_R=0}$, is tantamount to the statement that $|\psi\rangle$ belongs to a representation of $N_c$-ality opposite to that of the probe, or better, that it combines with the probe into a state of vanishing $N_c$-ality. This opens an interesting perspective: Within the Minkowskian formulation, it seems that one does not need the whole machinery of center transformations but only a remnant of it, in the form of the vanishing $N_c$-ality of the states formed by combining those states of definite net quark number that enter the trace and the probe. 

Of course, the fact that Eq.~(\ref{eq:conj}) leads to Eq.~(\ref{eq:lRt}) is not proof that (\ref{eq:conj}) holds true. A proof, if any, lies beyond the scope of this work. We believe, however, that \eqref{eq:conj} could shed some light on the status of center symmetry in the Hamiltonian framework, while establishing connections between $N_c$-ality, center-symmetry and confinement. To illustrate the usefulness of the picture relying on \eqref{eq:conj}, we now show how it can be used to derive the main result of this work regarding the behavior of $\ell_R$ at low temperatures and $\smash{|\mu|<M}$. Of course, once we gain access to the behavior of $\ell_R$, that of $\Delta F_R$ and $\Delta Q_R$ follows automatically.

\subsection{Low temperature behavior}
At low temperatures, the traces $Z_R$ and $Z$ are controlled by states that minimize  $\smash{H-\mu Q}$. Moreover, if $\smash{|\mu|<M}$, this involves only states with low energies. This means that the  combination $\smash{H-\mu Q}$ can be approximated as
\beq
H-\mu\,Q & = & M(N_q+N_{\bar q})-\mu(N_q-N_{\bar q})\nonumber\\
& = & (M-\mu)N_q+(M+\mu)N_{\bar q}\,,\label{eq:HmmuQ}
\eeq
where $N_q$ and $N_{\bar q}$ denote the number of quarks and antiquarks in each state. Here, we utilize the fact that quarks are heavy to approximate their energy by their mass and consider a non-relativistic limit in which $N_q$ and $N_{\bar q}$ make sense as separate quantities.

The minimization of $H-\mu\,Q$ needs to be done under the constraint (\ref{eq:conj}) which rewrites
\beq
N_q-N_{\bar q}=-\nu_R+N_ck\,.
\eeq
That is, we need to minimize
\beq
& & M(2N_q+\nu_R-N_c k)-\mu(-\nu_R+N_c k)\nonumber\\
& & \hspace{0.5cm}=\,2MN_q-N_ck(M+\mu)+\nu_R(M+\mu)\,,
\eeq
with respect to $N_q$ and $k$. Depending on how one proceeds, the discussion of $Z_R$ can be slightly different when $R$ has a vanishing $N_c$-ality or a non-vanishing one. A trick to treat these two cases simultaneously is to take $\smash{\nu_R=N_c}$ for the case of vanishing $N_c$-ality. This also allows one to treat the case of $Z$ together with that of $Z_R$.

With this in mind, we should now pay attention to the fact that, while $k\in\mathds{Z}$, $N_q$ is bounded from below by $-\nu_R+N_c k$ when $k\geq 1$ and by $0$ when $k\leq 0$. For a given $k\geq 1$, the minimal value of $H-\mu\,Q$ is then
\beq
(M-\mu)(-\nu_R+N_c k)\,,\label{eq:45}
\eeq
obtained for $\smash{N_q=-\nu_R+N_ck}$ and thus $\smash{N_{\bar q}=0}$, while for $k\leq 0$, it is
\beq
(M+\mu)(\nu_R-N_c k)\,,\label{eq:46}
\eeq
obtained for $\smash{N_q=0}$ and thus $\smash{N_{\bar q}=\nu_R-N_ck}$. We now need to compare all these possible values as we vary $k$. Since $\smash{|\mu|<M}$, the quantities (\ref{eq:45}) and (\ref{eq:46}) are minimized by $\smash{k=1}$ and $\smash{k=0}$ on their respective domains, and yield the values
\beq
H-\mu\,Q=(M-\mu)(-\nu_R+N_c)\,,\label{eq:48}
\eeq
corresponding to $\smash{N_q=-\nu_R+N_c}$, $\smash{N_{\bar q}=0}$ and thus $\smash{Q=-\nu_R+N_c}$, and
\beq
H-\mu\,Q=(M+\mu)\nu_R\,,\label{eq:49}
\eeq
corresponding to $\smash{N_q=0}$, $\smash{N_{\bar q}=\nu_R}$ and thus $\smash{Q=-\nu_R}$, respectively. Which of the two expressions (\ref{eq:48}) and (\ref{eq:49}) is minimal depends on whether $\smash{\mu>\mu_{\nu_R}}$ or  $\smash{\mu<\mu_{\nu_R}}$. 

We have thus found that, at low temperatures.
\beq
Z_R\propto\left\{
\begin{array}{cl}
e^{-\nu_R\beta(\mu+M)}\,, & \mbox{for } \mu<\mu_{\nu_R}\,,\\
e^{(N_c-\nu_R)\beta(\mu-M)}\,, & \mbox{for } \mu>\mu_{\nu_R}\,.
\end{array}
\right.\label{eq:ZRres}
\eeq
Taking $\smash{\nu_R=N_c}$ to deduced the behavior of $Z$, see the discussion above,  we find that $Z$ approaches a non-zero constant for any $\mu$ (such that $\smash{|\mu|<M}$). It then follows that, at low temperatures,
\beq
\ell_R\propto\left\{
\begin{array}{cl}
e^{-\nu_R\beta(\mu+M)}\,, & \mbox{for } \mu<\mu_{\nu_R}\,,\\
e^{(N_c-\nu_R)\beta(\mu-M)}\,, & \mbox{for } \mu>\mu_{\nu_R}\,,
\end{array}
\right.\label{eq:lRres}
\eeq
cf. \eqref{eq:fund}, which is the main result of this work. The proportionality symbol in (\ref{eq:ZRres}) and (\ref{eq:lRres}) comes from the fact that, despite $N_q$ and $N_{\bar q}$ being fixed (not only $Q$), the number of gluons is not fixed, and, then, various states may dominate the partition function with the same value of $Q$. The discussion of what happens at large $T$ or low $T$ with $\smash{|\mu|>M}$ is more involved since we cannot assume that the energies are small.

\subsection{Regarding center symmetry breaking}
The previous discussion also allows one to clarify the counterintuitive observation that center symmetry does not follow the order/disorder paradigm of other broken symmetries. In the usual paradigm, the symmetry breaking occurs because the ground state of the system is not invariant under a symmetry of the theory. Thus, the low-temperature, ordered phase is the broken phase, while in the high-temperature, disordered phase, thermal fluctuations restore the symmetry.

It is not easy to apply this logic to center symmetry because the latter does not really act on the states of the Minkowskian formulation. As we have seen, however, what plays the role of a proxy for center symmetry in this framework is the constraint (\ref{eq:conj}) on the states. We have conjectured that this constraint applies to all states involved in a given trace. Thus, in this sense, the ``center symmetry'' is always manifest at the level of the states. However, and as opposed to other symmetries, it only manifests in thermal averages at low temperatures where the minimization of $H-\mu Q$ select states with a definite value of $Q=-\nu_R$ or $Q=N_c-\nu_R$. In this case, the thermal average of $Q$ reflects directly the constraint on the states and thus ``center symmetry''. As one increases the temperature, however, thermal fluctuations bring states with other values of $Q=-\nu_R+N_c k$ with arbitrary $k$ and thus the average of $Q$ does not reflect anymore the constraint on the states

The manifestation of the constraints in thermal averages at low temperatures is what eventually ensures that the Polyakov loops of non-vanishing $N_c$-ality vanish in the low-temperature limit. Note that, strictly speaking, as long as the quarks have a finite mass, center symmetry is explicitly broken in the Euclidean functional integral formulation due to the boundary conditions of the quarks. However, in the low temperature limit, the boundary conditions do not matter, and so center symmetry is restored, which makes the Polyakov loops of non-vanishing $N_c$-ality vanish in the low temperature limit. So in both formulations, we obtain a consistent picture in the low temperature limit. In one case, it has to do with the restoration of center symmetry through a loss of memory of the boundary conditions in the low temperature limit, while in the other case, it has to do with the presence of a constraint on the states combined with a minimization principle.

The actual spontaneous breaking of symmetry occurs in the pure glue theory, corresponding to the limit where the masses of the quarks in the bath are taken to infinity. Our interpretation also allows one to clarify the specifics of the breaking. For large but finite quark masses, it is always possible to find quarks and antiquarks from the bath that will compensate for the $N_c$-ality of the probe, so the free energy should always be finite. However, as the masses of the quarks increase, it should become more and more difficult to find those quarks and antiquarks in the bath since their probabilities decrease exponentially. We then expect the free energy to grow and the Polyakov loop to decrease. If no singularity is present, sending the masses of the quarks in the bath to infinity in the presence of an infinitely heavy probe should commute with bringing an infintely massive probe into a purely gluonic bath. However, in this latter case, there are no quarks and antiquarks in the bath to compensate for the $N_c$-ality of the probe (if the probe has a non-vanishing $N_c$-ality), the partition function $Z_R$ (and thus the Polaykov loop $\ell_R$) vanishes, and the free energy is infinite. This case should occur in some interval of temperature containing the $\smash{T=0}$ since we have seen that, in the presence of finite quark masses, the Polyakov loop approaches $0$ in the zero-temperature limit.

Above some temperature, however, it may happen that sending the mass of the quarks in the bath to infinity in the presence of an infinitely heavy probe does not commute with bringing the infinitely massive probe into a purely gluonic medium. In that case, the system keeps a memory of the presence of quarks, and the Polyakov loops never reach $0$. The same occurs within the Euclidean functional integral formulation, where the system keeps memory of the quark boundary conditions despite sending their mass to infinity and so (\ref{eq:lRt}) does not imply anymore that $\ell_R$ vanishes in the pure glue limit.

\section{Conclusions}
We have investigated how the addition of a probe in a bath of SU($N_c$) quarks and gluons at finite temperature and chemical potential affects the net quark number of the bath. We have argued that this depends exclusively on the $N_c$-ality $\nu_R$ of the color representation $R$ of the probe. At low temperatures and for $\mu_{\nu_R}<\mu<M$, the probe is absorbed into a state with the same net quark number as one baryon, whereas for $-M<\mu<\mu_{\mu_R}$, the probe is absorbed into a state with the same net quark number as a bunch of mesons. These findings generalize those obtained in Ref.~\cite{MariSurkau:2025how} in the SU(3) case.

Our results are based on a number of assumptions, some of which are well-established properties of non-Abelian gauge theories. Some others are shared by most, if not all, successful models for the confinement/deconfinement transition. Testing quark number gains in lattice simulations in heavy-quark QCD could allow one to shed some light on which of these properties are fundamental. We plan to investigate this in the future.

We have also proposed an interpretation of our results from a Minkowskian perspective. Even though this interpretation remains conjectural, and requires a proper proof that we plan to investigate in the future, it highlights the role of $N_c$-ality as a proxy for center symmetry in the Minkowskian formulation. This interpretation could also allows one to clarify some of the specifics of center symmetry, in particular, the fact that it does not fit the usual order/disorder paradigm of other broken symmetries.

On a broader perspective, net quark number gains are just a particular example of observables that could allow one to probe the degrees of freedom in the various phases of QCD. More generally, one could study the expectation value of other conserved charges in the presence of an external probe.

\appendix

\section{Polyakov loop potential}\label{app:pot}

Following similar techniques as in Ref.~\cite{Dumitru:2013xna}, one writes
\beq
V_{\rm quark}(\ell_\nu)=W_{\rm quark}(r^j)\,,
\eeq
with (a summation over $j$ is implied)
\beq
\ell_\nu=\frac{1}{C_{N_c}^\nu}\sum_{\rho_\nu} e^{i\rho_\nu^j r^j}\,,\label{eq:lr}
\eeq
and
\beq
W_{\rm quark}(r^j) & = & -\frac{T}{\pi^2}\sum_\rho\int_0^\infty \!\!dq\,q^2\nonumber\\
& \times & \Big\{\ln\Big[1+e^{-\beta\,(\varepsilon_q-\mu)+i r^j\rho^j)}\Big]\nonumber\\
& & \hspace{0.2cm}+\,\ln\Big[1+e^{-\beta\,(\varepsilon_q+\mu)-i r^j\rho^j)}\Big]\Big\}.\label{eq:Wr}
\eeq
It should be stressed that the weights $\rho_\nu$ in Eq.~(\ref{eq:lr}) are those of the fundamental representation $\nu$, while those in Eq.~(\ref{eq:Wr}) correspond to $\smash{\nu=1}$. In fact, the former can be obtained as all possible sums of the latter, involving $\nu$ terms, and without repetitions:
\beq
\rho_{(k_1)}^j+\cdots+\rho_{(k_\nu)}^j\,,
\eeq
where we have used the subscript $(k)$ to label the various weights of the fundamental representation $\smash{\nu=1}$ and we take $k_1<\cdots<k_\nu$. Let us now use this relation between the weights of the fundamental representation $\nu$ and those of the fundamental representation $\smash{\nu=1}$ can be used to deduce an explicit expression for $V_{\rm quark}(\ell_\nu)$.

Combining the logarithms in Eq.~(\ref{eq:Wr}), one arrives at
\begin{eqnarray}
& &  \ln\prod_\rho\Big[1+e^{-\beta\,(\varepsilon_q\mp\mu)\pm ir^j\rho^j}\Big]\label{eq:ln}\\
& & \hspace{0.1cm}=\,\ln \Big[1+\sum_{\nu=1}^{N_c}e^{-\nu\beta\,(\varepsilon_q\mp\mu)}\!\!\!\!\!\sum_{k_1<\cdots<k_\nu}e^{\pm ir^j\big(\rho_{(k_1)}^j+\cdots+\rho_{(k_\nu)}^j\big)}  \Big],\nonumber
\end{eqnarray}
where we have used the subscript $(k)$ to label the various weights of the fundamental representation $\smash{\nu=1}$. We recognize the weights of the fundamental representation $\nu$ in each of the exponents, and the sum over $\smash{k_1<\cdots<k_\nu}$ stands for the sum over all such weight. We can then use (\ref{eq:lr}) to replace this sum by $C_{N_c}^\nu\ell_\nu$ in the case where the exponentials feature a $+$. In the case where the exponential feature a minus, we use that the weights of the fundamental representation $\smash{\nu=1}$ add up to $0$:\footnote{Recall also that any choice of $N_c-1$ weights out of the $N_c$ forms a linearly independent set.}
\beq
\sum_{k=1}^{N_c}\rho_{(k)}=0\,.\label{eq:sum0}
\eeq
Then, we can write
\beq
& & \sum_{k_1<\cdots<k_\nu}e^{- ir^j\big(\rho_{(k_1)}^j+\cdots+\rho_{(k_\nu)}^j\big)}\nonumber\\
& & \hspace{1.0cm}=\!\!\!\sum_{k_1<\cdots<k_{N_c-\nu}}e^{ir^j\big(\rho_{(k_1)}^j+\cdots+\rho_{(k_{N_c-\nu})}^j\big)}\,,
\eeq
which can be replaced by $C_{N_c}^\nu\ell_{N_c-\nu}$. Upon making these replacements, we arrive at the formula (\ref{eq:Vquark}) given in the main text.

Center transformations can be seen to correspond to transformations of the Euclidean gauge field $A_\mu^a t_R^a$ in(\ref{eq:lR}) of the form
\beq
A_\mu^a t_R^a\to {\cal U}_R A_\mu^a t_R^a\,{\cal U}_R^{-1}+i\,{\cal U}_R\partial_\mu {\cal U}_R^\dagger\,,
\eeq
with
\beq
{\cal U}_R(\tau)=e^{i\frac{\tau}{\beta}4\pi \rho_{(k)}^jt^j_R}\,,
\eeq
see Ref.~\cite{vanEgmond:2023lfu}, where $\rho_{(k)}$ denotes any of the weights of the fundamental representation $\smash{\nu=1}$. Under these transformations the traced path-ordered exponential in (\ref{eq:lR}) transforms as
\beq
& & {\rm tr}\,{\cal P}\,e^{ig\int_0^\beta d\tau\,A_4^a(\tau,\vec{x})t^a_R}\nonumber\\
& & \hspace{0.5cm}\to {\rm tr}\,{\cal U}_R^\dagger(0)\,{\cal U}_R(\beta){\cal P}\,e^{ig\int_0^\beta d\tau\,A_4^a(\tau,\vec{x})t^a_R}\,,
\eeq
with
\beq
{\cal U}_R^\dagger(0)\,{\cal U}_R(\beta)=e^{i4\pi \rho_{(k)}^jt^j_R}\,.
\eeq
To continue, it is convenient to decompose ${\cal U}_R^\dagger(0)\,{\cal U}_R(\beta)$, in the basis $|\rho_R\rangle$ that diagonalizes simultaneously all the $t_R^j$:
\beq
{\cal U}_R^\dagger(0)\,{\cal U}_R(\beta) & = & e^{i4\pi \rho_{(k)}^jt^j_R}\sum_{\rho_R}|\rho_R\rangle\langle\rho_R|\nonumber\\
& = & \sum_{\rho_R}e^{i4\pi \rho_{(k)}^j\rho_R^j}|\rho_R\rangle\langle\rho_R|\,.\label{eq:int}
\eeq
Now, according to the Young tableau of $R$, any weight of $R$ obeys
\beq
\rho_R=\sum_{k=1}^{N_c}n_k \rho_{(k)}\,,\label{eq:rhon}
\eeq
with $\smash{n_k\in\mathds{N}}$ and where $\sum_{k=1}^{N_c} n_k$ is the total number of elements of the tableau, and thus equals the $N_c$-ality $\nu_R$ of $R$, modulo $N_c$. Using that
\beq
\rho_{(k)}^j\rho_{(l)}^j=\frac{1}{2}\left(\delta_{k\ell}-\frac{1}{N_c}\right),
\eeq
it then follows that
\beq
\rho_{(k)}^j\rho_R^j & = & \frac{n_k}{2}-\frac{\sum_{k=1}^{N_c}n_k}{2N_c}\nonumber\\
& = & -\frac{\nu_R}{2N_c}\,{\rm mod}\,\frac{1}{2}\,.
\eeq
Using this in Eq.~(\ref{eq:int}), we arrive at
\beq
{\cal U}_R^\dagger(0)\,{\cal U}_R(\beta)=e^{-i2\pi\nu_R/N_c}\mathds{1}
\eeq
This implies that, under a center transformation, the Polyakov loop (\ref{eq:lR}) gets multiplied by a factor $e^{-i2\pi\nu_R/N_c}$.\footnote{A similar conclusion can be arrived at using the tree-level Polyakov loops in terms of their associated weights.} In the presence of quarks, one should not forget that the quark boundary conditions are also affected.

Let us finally gather some additional properties on the weights $\rho_R$ of a representation $R$. First, the property (\ref{eq:sum0}) extends to the weights of any representation. Let us start noticing that the integer $n_k$ in (\ref{eq:rhon}) depends on the chosen weight $\rho_R$ in the representation $R$. However, when summing over possible $\rho_R$ of that representation, and because the various $\rho_{(k)}$ are equivalent, the sum $\sum_{\rho_R}n_k(\rho_R)$ should not depend on $k$. We then have
\beq
\sum_{\rho_R}\rho_R & = & \sum_{k=1}^{N_c}\Big(\sum_{\rho_R}n_k(\rho_R)\Big)\rho_{(k)}\nonumber\\
& = & \sum_{\rho_R}n_1(\rho_R)\times\sum_{k=1}^{N_c}\rho_{(k)}=0\,.
\eeq
Second, note that one may rewrite $\rho_R$ in many equivalent forms:
\beq
\rho_R=\sum_{k=1}^{N_c}m_k \rho_{(k)}
\eeq
with $\smash{m_h\in\mathds{Z}}$. Combining this with Eq.~(\ref{eq:rhon}), one finds
\beq
0=\sum_{k=1}^{N_c}(m_k-n_k) \rho_{(k)}\,.\label{eq:wd}
\eeq
Because of (\ref{eq:sum0}) and since any subset of $N_c-1$ weights $\rho_{(k)}$ out of the $N_c$ is linearly independent, the only way that (\ref{eq:wd}) can be true is if $m_k-n_k$ is an integer independent of $k$. Summing over $k$, one then finds that
\beq
\sum_{k=1}^{N_c}m_k=\sum_{k=1}^{N_c}n_k\,{\rm mod}\,N_c\,,
\eeq
and so $\sum_k m_k$ also equals the $N_c$-ality of $R$, modulo $N_c$.

\section{Product of Polyakov loops}\label{app:ordering}

Consider a product $\ell_{\nu_1}\cdots\ell_{\nu_p}$ of fundamental Polyakov loops and define $\nu$ as the remainder in the Euclidean division of $\nu_1+\cdots+\nu_p$ by $N_c$:
\beq
\nu_1+\cdots+\nu_p=\nu+N_ck\,.
\eeq
We would like to show that the single fundamental Polyakov loop $\ell_\nu$ always dominates over the product $\ell_{\nu_1}\cdots\ell_{\nu_p}$ in the low temperature limit and for $\smash{|\mu|<M}$. To this purpose we will evaluate the ratio $\ell_{\nu_1}\cdots\ell_{\nu_p}/\ell_\nu$ over the various intervals of $\mu$. We should pay attention to the fact that, unlike the $\nu_i$ which are comprised between $1$ and $N_c-1$, $\nu$ is comprised between $0$ and $N_c-1$. When $\smash{\nu=0}$, we then need to specify what we mean by $\ell_\nu$. We set $\smash{\ell_0=1}$. 

To avoid cluttering our formulas, we introduce the following bracket notation
\beq
e^{a\beta\mu+b\beta M}(\beta M)^{3c/2}\equiv [a,b,c]\,.
\eeq
 In particular the bracket associated to $\ell_\nu$ is:
\beq
\ell_\nu:\left\{
\begin{array}{cl}
[-\nu,-\nu,-1]\,, & \mbox{for } \mu<\mu_\nu\,,\\
{[N_c-\nu,\nu-N_c,-1]\,,} & \mbox{for } \mu>\mu_\nu\,.
\end{array}
\right.
\eeq
A special treatment needs to be done when $\smash{\nu=0}$ since $\smash{\mu_0=M}$ the associated bracket is
\beq
\ell_0:[0,0,0] \mbox{ for } \mu<\mu_0\,.
\eeq
Products and ratios of such exponential/power-law expressions translate into additions and subtractions of the bracket.

We consider first the case $\smash{p=2}$, and assume without loss of generality that $\smash{\nu_1\leq\nu_2}$. Then $k$ can equal $0$ or $1$. In the first case, $\smash{\nu=\nu_1+\nu_2>\nu_2\geq\nu_1}$, and thus $\mu_{\nu}<\mu_{\nu_2}\leq\mu_{\nu_1}$. Note also that $\nu$ cannot vanish in this case. In the right-most interval of $\mu$ (and below $M$), the bracket associated to $\ell_{\nu_1}\ell_{\nu_2}/\ell_\nu$ writes
\beq
& & [N_c-\nu_1,\nu_1-N_c,-1]+[N_c-\nu_2,\nu_2-N_c,-1]\nonumber\\
& & -\,[N_c-\nu_1-\nu_2,\nu_1+\nu_2-N_c,-1]\nonumber\\
& & =\,[N_c,-N_c,-1]\,,\label{eq:b1}
\eeq
which is exponentially suppressed at low temperatures over the considered interval of $\mu$. To continue, let us momentarily assume that $\nu_1\neq\nu_2$. Then, crossing to the next interval, that is crossing $\mu_{\nu_1}$ is tantamount to adding $[-N_c,N_c-2\nu_1,0]$ to the previously obtained bracket. We then find
\beq
[0,-2\nu_1,-1]\,,
\eeq
which is still exponentially suppressed over the considered interval. Crossing to the next interval, we now need to add $[-N_c,N_c-2\nu_2,0]$ thus yielding
\beq
[-N_c,N_c-2(\nu_1+\nu_2),-1]\,,\label{eq:b3}
\eeq
which is again easily checked to be exponentially suppressed over the considered interval. The only difference in the case $\smash{\nu_1=\nu_2}$ is that one goes directly from bracket (\ref{eq:b1}) to bracket (\ref{eq:b3}). We can now lift the assumption $\smash{\nu_1<\nu_2}$. Finally, crossing to the last interval, we need to subtract $[-N_c,N_c-2\nu,0]$ and we find
\beq
[0,0,-1]\,,
\eeq
which is polynomially suppressed.

A similar reasonning applies when $\smash{k=1}$, in which case $\nu=\nu_1+\nu_2-N_c<\nu_1\leq\nu_2$, and thus $\smash{\mu_{\nu_2}\leq\mu_{\nu_1}<\mu_\nu}$. In this case, $\nu$ can vanish but let us first assume that it does not. Over the rightmost interval (and below $M$), we find
\beq
& & [N_c-\nu_1,\nu_1-N_c,-1]+[N_c-\nu_2,\nu_2-N_c,-1]\nonumber\\
& & -\,[2N_c-\nu_1-\nu_2,\nu_1+\nu_2-2N_c,-1]\nonumber\\
& & =\,[0,0,-1]\,,
\eeq
which is polynomially suppressed. Over the subsequent intervals, we find
\beq
[N_c,-3N_c+2(\nu_1+\nu_2),-1]\,,
\eeq
\beq
[0,-2N_c+2\nu_2,-1]\,,
\eeq
and
\beq
[-N_c,N_c,-1]\,,
\eeq
which are all exponentially suppressed over the considered intervals. In the case where $\smash{\nu=0}$, there is one less interval to consider and the subsequent brackets are
\beq
[2N_c-\nu_1-\nu_2,\nu_1+\nu_2-2N_c,-2]\,,
\eeq
\beq
[N_c-\nu_1-\nu_2,-\nu_1+\nu_2-N_c,-2]\,,
\eeq
and
\beq
[-\nu_1-\nu_2,-\nu_1-\nu_2,-2]\,,
\eeq
which are all exponentially suppressed. We have implicitly assumed that $\smash{\nu_1\neq\nu_2}$. If $\smash{\nu_1=\nu_2}$, some of the intermediate brackets are missing but the conclusion is the same. In all cases, we find that $\ell_{\nu_1}\ell_{\nu_2}$ is suppressed over $\ell_\nu$ at low temperatures and $\smash{|\mu|<M}$.

To extend this result to an arbitrary product of fundamental Polyakov loops, we can now proceed recursively. Suppose that we have shown that any product $\ell_{\nu_1}\cdots\ell_{\nu_p}$ of $p$ fundamental Polyakov loops is suppressed with respect to the corresponding $\ell_\nu$ and consider another product $\ell_{\nu_1}\cdots\ell_{\nu_p}\ell_{\nu_{p+1}}$ of $p+1$ fundamental Polyakov loops. If the corresponding $\nu$ vanishes, it is clear that the product is suppressed with respect to $\ell_\nu$, so let us assume that it does not. Without loss of generality, we can also assume that the $\nu$ associated to $\nu_1,\cdots\nu_p$ does not vanish either. We denote it by $\sigma$. We have
\beq
\nu_1+\cdots+\nu_p=\sigma+N_c h\,,
\eeq
and necessarily
\beq
\sigma+\nu_{p+1}=\nu+N_c j\,,
\eeq
such that
\beq
\nu_1+\cdots+\nu_p+\nu_{p+1}=\nu+N_c k\,,
\eeq
with $\smash{k=h+j}$. We now write
\beq
\frac{\ell_{\nu_1}\cdots\ell_{\nu_p}\ell_{\nu_{p+1}}}{\ell_\nu}=\frac{\ell_{\nu_1}\cdots\ell_{\nu_p}}{\ell_\sigma}\frac{\ell_\sigma\ell_{\nu_{p_1}}}{\ell_\nu}\,.
\eeq
Each factor in the right-hand side being suppressed at low temperatures and for any $\smash{|\mu|<M}$, so is the same for the ratio in the left-hand side. We have thus shown that the result extends to a product of $p+1$ fundamental Polyakov loops, and thus, by recursion, it applies to any finite product of fundamental Polyakov loops.

\section{Beyond the linear approximation}\label{app:linear}

In the main text, we obtained the following behavior for the fundamental Polyakov loops at low temperature and $\smash{|\mu/M|<1}$:
\beq
\ell_\nu\propto e^{(N_c-\nu)\beta\mu}f_{(N_c-\nu)\beta M}\,,\label{eq:asymp1}
\eeq
for $\smash{1-2\nu/N_c<\mu/M<1}$, and
\beq
\ell_\nu\propto e^{-\nu\beta\mu}f_{\nu\beta M}\,,\label{eq:asymp2}
\eeq
for $\smash{-1<\mu/M<1-2\nu/N_c}$. These behaviors were obtained from a linear approximation of the equations determining the fundamental Polyakov loops around the confining point at which the fundamental Polyakov loops vanish. Although this approximation is justified at asymptotically small temperatures, it is not adapted to small but finite temperatures. The reason is that higher-order terms in the Taylor expansion can be of the same order as the 
linear terms when the asymptotic expressions (\ref{eq:asymp1}) or (\ref{eq:asymp2}) are used.

Here, we go beyond the linear approximation by analyzing the role of higher-order terms in the expansion. To make the discussion tractable, however, we restrict to the interval\footnote{Other intervals could be treated in a similar fashion, but the analysis becomes more involved.} $1-2/N_c<\mu/M<1$ and we proceed to a consistency check rather than a full rigorous proof. Namely, by assuming the asymptotic expressions (\ref{eq:asymp1}) and (\ref{eq:asymp2}), we retain in the equations all contributions that are of the same order as the linear order. The system of equations then simplifies into a triangular system, which can then be explicitly solved. In a final step, we check that the solution of this system is compatible with the assumed asymptotic behaviors, with, however, modified pre-factors as compared to those obtained using the linear approximation.

\subsection{Equations in the interval $1-2/N_c<\mu/M<1$}
Suppose then that we expand the equation (\ref{eq:gap}) around the confining point. We write the left-hand side of this equation as
\beq
\frac{\partial V_{\rm glue}}{\partial\ell_\nu}=\sum_{p\geq 1}\frac{1}{p!}\frac{\partial^{p+1} V_{\rm glue}}{\partial\ell_\nu\partial\ell_{\nu_1}\cdots\partial\ell_{\nu_p}}\ell_{\nu_1}\cdots\ell_{\nu_p}\,,\label{eq:Taylor}
\eeq
where the derivatives in the right-hand side need to be understood as taken at the confining point. 

We shall first assume that $\mu/M$ is taken in the interval 
\beq
1-\frac{2}{N_c}<\frac{\mu}{M}<1\,,
\eeq
in which case, we can assume that all fundamental Polyakov loops scale as in Eq.~(\ref{eq:asymp1}). This implies that each term of the sum (\ref{eq:Taylor}) scales as
\beq
e^{(N_cp-\nu_1\cdots-\nu_p)\beta(\mu-M)}\,.\label{eq:factor}
\eeq
Now, the center symmetry encoded in Eq.~(\ref{eq:center}), imposes that $\partial^{p+1}V_{\rm glue}/\partial\ell_\nu\partial\ell_{\nu_1}\cdots\partial\ell_{\nu_p}$ vanishes if $\nu+\nu_1+\cdots+\nu_p$ is not a multiplie of $N_c$. We can thus assume that, for each term in the sum (\ref{eq:Taylor}), we have
\beq
\nu+\nu_1+\cdots+\nu_p=N_ck\,,
\eeq
for some integer $k\leq p$. Then, the exponential factor (\ref{eq:factor}) rewrites
\beq
e^{(N_c (p-k)+\nu)\beta(\mu-M)}\,.\label{eq:scale}
\eeq
This means that, for each value of $p$ in the sum (\ref{eq:Taylor}), we can classify the various products of $p$ fundamental Polyakov loops according to the associated value of $k$: the higher the value of $k$, the less the product of Polyakov loops is suppressed at low temperatures.

Now, since each $\nu_i$ lies between $1$ and $N_c-1$, $N_ck$ is bounded by $\nu-p+N_cp$. This implies on the one hand that $k$ is strictly less than $p$ whenever $\smash{p>\nu}$. On the other hand, for $\smash{p\leq\nu}$, there is always a choice of the $\nu_i$ such that $\smash{k=p}$ and thus dominates over any term in the Taylor expansion with $\smash{p>\nu}$. Take first $\smash{p=\nu}$. Then the only possible choice with $\smash{k=p}$ is the one for which all $\nu_i$ equal $N_c-1$. If we decrease $p$ by one unit, we can still ensure that $\smash{k=p}$ by demanding that one of the $\nu_i$ now equals $\smash{N_c-2}$. As we decrease $p$ by one further unit, we can now demand that this same $\nu_i$ now equals $N_c-3$ or that another $\nu_i$ equals $N_c-2$. We continue that way until we reach $\smash{p=1}$, in which case the only choice is that $\nu_1$ equals $N_c-\nu$.

What we learn from this analysis is, first, that the Taylor expansion can be safely truncated to order $\smash{p=\nu}$. The equations fixing the fundamental Polyakov loops then rewrite as
\beq
& & \sum_{p=1}^\nu\frac{1}{p!}\frac{\partial^{p+1} V_{\rm glue}}{\partial\ell_\nu\partial\ell_{\nu_1}\cdots\partial\ell_{\nu_p}}\ell_{\nu_1}\cdots\ell_{\nu_p}\nonumber\\
& & \hspace{0.5cm}\simeq\, \alpha_\nu\Big(e^{\nu\beta\mu} f_{\nu\beta M}+e^{-(N_c-\nu)\beta\mu} f_{(N_c-\nu)\beta M}\Big)\,,\label{eq:eqs}\nonumber\\
\eeq
with
\beq\label{eq: center sym dominant}
\nu+\nu_1+\cdots+\nu_p=N_cp\,.
\eeq
Second, we know that for each $\smash{p\leq\nu}$, the $\nu_i$ are bounded from below by $N_c-(\nu-p+1)$. This implies that the equations (\ref{eq:eqs}) form a triangular system that can be solved iteratively starting from $\ell_{N_c-1}$, as we now discuss.

\subsection{Solution in the interval $1-2/N_c<\mu/M<1$}
In what follows, it will be convenient to rewrite the triangular system as
\beq
& & \sum_{p=1}^\nu\frac{1}{p!}X_{\nu_1\cdots\nu_p}\ell_{N_c-\nu_1}\cdots\ell_{N_c-\nu_p}\nonumber\\
& & \hspace{0.5cm}\simeq\, \alpha_\nu\Big(e^{\nu\beta\mu} f_{\nu\beta M}+e^{-(N_c-\nu)\beta\mu} f_{(N_c-\nu)\beta M}\Big)\,,\label{eq:eqs2}\nonumber\\
\eeq
with $\nu_1+\cdots+\nu_p=\nu$ and $\smash{\nu_i\leq\nu-p+1}$. We have also introduced the shorthand notation
\beq
X_{\nu_1\cdots\nu_p}\equiv\frac{\partial^{p+1} V_{\rm glue}}{\partial\ell_\nu\partial\ell_{N_c-\nu_1}\cdots\partial\ell_{N_c-\nu_p}}\,,
\eeq
to avoid cluttering.

In the considered interval, the first term in the right-hand side of Eq.~(\ref{eq:eqs2}) dominates, and so the triangular system becomes
\beq
\sum_{p=1}^\nu\frac{1}{p!}X_{\nu_1\cdots\nu_p}\ell_{N_c-\nu_1}\cdots\ell_{N_c-\nu_p}\simeq\, \alpha_\nu e^{\nu\beta\mu} f_{\nu\beta M}\,.
\eeq
The first equation is obtained for $\smash{\nu=1}$:
\beq
X_1\ell_{N_c-1}\simeq \alpha_1 e^{\beta\mu} f_{\beta M}\,.
\eeq
It fixes $\ell_{N_c-1}$ to the same expression as the one obtained in the main text using the linear approximation
\beq
\ell_{N_c-1}\simeq\gamma_1\, e^{\beta\mu}f_{\beta M}\,,\label{eq:g1}
\eeq
with 
\beq
\gamma_1\equiv\frac{\alpha_1}{X_1}\,.
\eeq
The next equation is
\beq
X_2\ell_{N_c-2}+\frac{1}{2}X_{11}\ell_{N_c-1}^2\simeq\alpha_2 e^{2\beta\mu} f_{2\beta M}\,,
\eeq
and fixes $\ell_{N_c-2}$ to
\beq
\ell_{N_c-2}\simeq \gamma_2\,e^{2\beta\mu}f_{2\beta M}\,,
\eeq
with
\beq
\gamma_2=\frac{\alpha_2}{X_2}-\frac{\gamma_1^2}{2}\frac{X_{11}}{X_2}\frac{f^2_{\beta M}}{f_{2\beta M}}\,,
\eeq
and where we have made use of Eq.~(\ref{eq:g1}). Thus, the exponential suppression of $\ell_{N_c-2}$, including its $\mu$-dependence, is the same as the one obtained using the linear approximation, up to a modified pre-factor. In fact, this modified pre-factor deviates from the naive by a power-law suppressed contribution.

Suppose now that we have shown that 
\beq
\ell_{N_c-\nu}\simeq\gamma_\nu\, e^{\nu\beta M}f_{\nu\beta M}\,,\label{eq:C}
\eeq
for any $\smash{\nu\leq \nu_0-1}$. Consider the equation for $\ell_{N_c-\nu_0}$. It reads
\beq
& & X_{\nu_0}\ell_{N_c-\nu_0}\\
& & +\,\sum_{p=2}^{\nu_0}\frac{1}{p!}X_{\nu_1\cdots\nu_p}\ell_{N_c-\nu_1}\cdots\ell_{N_c-\nu_p}\simeq\alpha_{\nu_0} e^{\nu_0\beta\mu} f_{\nu_0\beta M}\,,\nonumber
\eeq
with $\nu_1+\cdots+\nu_p=\nu_0$ and $\nu_i\leq \nu_0-p+1$. For those terms with $p\geq 2$, we can use (\ref{eq:C}) and thus
\beq
\ell_{N_c-\nu_0}\simeq\gamma_{\nu_0}\, e^{\nu_0\beta M}f_{\nu_0\beta M}\,,
\eeq
with
\beq
\gamma_{\nu_0}=\frac{\alpha_{\nu_0}}{X_{\nu_0}}-\sum_{p=2}^{\nu_0}\frac{\gamma_{\nu_1}\cdots\gamma_{\nu_p}}{p!}\frac{X_{\nu_1\cdots\nu_p}}{X_{\nu_0}}\frac{f_{\nu_1\beta M}\cdots f_{\nu_p\beta M}}{f_{\nu_0\beta M}}\,.\nonumber\\
\eeq

\section{Polyakov loops and character identities}\label{app:character}

A representation $R$ of SU($N_c$) associates to each element $\smash{U\in\,\,}$SU($N_c$), an invertible linear map $R(U)$ over a vector space $V$ over $\mathds{C}$, which preserves the group law: $R(U_1U_2)=R(U_1)\circ R(U_2)$. The character of $R$ is a function $\chi_R$ from SU($N_c$) to $\mathds{C}$ defined as
\beq
\chi_R(U)={\rm tr}\,R(U)\,.
\eeq
We note that $\chi_R(\mathds{1})$ is the dimension $d_R$ of the representation.

Given two representations $R_1$ and $R_2$, one can form their tensor product $\smash{R_1\otimes R_2}$. It associates to $U$ the linear map
\beq
(R_1\otimes R_2)(U)\equiv R_1(U)\otimes R_2(U)\,,
\eeq
over $V_1\otimes V_2$. One can also form the direct sum $R_1\oplus R_2$. It associates to $U$ the linear map
\beq
(R_1\oplus R_2)(U)\equiv R_1(U)P_1+R_2(U)P_2\,,
\eeq
over $V_1\times V_2$, where $P_i$ denotes the projection from $V_1\times V_2$ to $V_i$. The characters of $R_1$, $R_2$, $R_1\otimes R_2$ and $R_1\oplus R_2$ are related by
\beq
\chi_{R_1\otimes R_2} & = & \chi_{R_1}\chi_{R_2}\,,\label{eq:firstc}\\
\chi_{R_1\oplus R_2} & = & \chi_{R_1}+\chi_{R_2}\,.\label{eq:secondc}
\eeq
Evaluating these identities for $\smash{U=\mathds{1}}$, we find in particular $\smash{d_{R_1\otimes R_2}=d_{R_1}d_{R_2}}$ and $\smash{d_{R_1\oplus R_2}=d_{R_1}+d_{R_2}}$.

Consider now an element of the group of the form
\beq
U=e^{i\theta^at^a}\,,
\eeq
where the $t^a$ are the generators of the group. For a given representation $R$, we have
\beq
R(e^{i\theta^at^a})=e^{i\theta^at^a_R}\,,
\eeq
where the $t^a_R$ are known as the generators of the representation $R$. The above identities then write
\beq
e^{i\theta^at^a_{R_1\otimes R_2}} & = & e^{i\theta^at^a_{R_1}}\otimes e^{i\theta^at^a_{R_2}}\,,\\
e^{i\theta^at^a_{R_1\oplus R_2}} & = & e^{i\theta^at^a_{R_1}}P_1+ e^{i\theta^at^a_{R_2}}P_2\,,
\eeq
and
\beq
{\rm tr}\,e^{i\theta^at^a_{R_1\otimes R_2}} & = & {\rm tr}\,e^{i\theta^at^a_{R_1}}\,{\rm tr}\, e^{i\theta^at^a_{R_2}}\,,\label{eq:p}\\
{\rm tr}\,e^{i\theta^at^a_{R_1\oplus R_2}} & = & {\rm tr}\,e^{i\theta^at^a_{R_1}}+ {\rm tr}\,e^{i\theta^at^a_{R_2}}\,.\label{eq:s}
\eeq

It can now be argued that the previous identities are valid when 
\beq
\theta^a=\int_0^\beta d\tau\,A_0^a(\tau,\vec{x})\,,
\eeq
and in the presence of a time-ordering operator in front of each of the exponentials. Upon taking the average in the presence of $e^{-S_{\rm QCD}}$, one then deduces that
\beq
d_{R_1\oplus R_2}\ell_{R_1\oplus R_2}=d_{R_1}\ell_{R_1}+d_{R_2}\ell_{R_2}
\eeq
while
\beq
& & d_{R_1\otimes R_2}\ell_{R_1\otimes R_2}\\
& & \hspace{0.3cm}=\left\langle{\rm tr}\,{\cal P}\,e^{i\int_0^\beta d\tau\,A_0^a(\tau,\vec{x})t^a_{R_1}}{\rm tr}\,{\cal P}\,e^{i\int_0^\beta d\tau\,A_0^a(\tau,\vec{x})t^a_{R_2}}\right\rangle\,.\nonumber
\eeq
We stress that the right-hand side of this second identity has no reason to equal $d_{R_1}d_{R_2}\ell_1\ell_2$, and then the exact Polyakov loops do not fulfill the first character identity (\ref{eq:firstc}). This implies that, when $R_1\otimes R_2$ decomposes as
\beq
R_1\otimes R_2={\bigoplus}_iR'_i\,,
\eeq
we do not have in general
\beq
d_{R_1}d_{R_2}\ell_{R_1}\ell_{R_2}={\sum}_i d_{R'_i}\ell_{R'_i}\,,\label{eq:valid}
\eeq
but only
\beq
d_{R_1}d_{R_2}\ell_{R_1\otimes R_2}={\sum}_i d_{R'_i}\ell_{R'_i}\,.
\eeq
Equation (\ref{eq:valid}) is usually valid at tree-level order of a semi-classical expansion in a given gauge. This is because, in this case,
\beq
& & \left\langle{\rm tr}\,{\cal P}\,e^{i\int_0^\beta d\tau\,A_0^a(\tau,\vec{x})t^a_{R_1}}{\rm tr}\,{\cal P}\,e^{i\int_0^\beta d\tau\,A_0^a(\tau,\vec{x})t^a_{R_2}}\right\rangle\nonumber\\
& & \hspace{1.0cm}\simeq\,{\rm tr}\,{\cal P}\,e^{i\int_0^\beta d\tau\,\langle A_0^a(\tau,\vec{x})\rangle t^a_{R_1}}\,{\rm tr}\,{\cal P}\,e^{i\int_0^\beta d\tau\,\langle A_0^a(\tau,\vec{x})\rangle t^a_{R_2}}\nonumber\\
& & \hspace{1.0cm}\simeq\,\left\langle{\rm tr}\,{\cal P}\,e^{i\int_0^\beta d\tau\,A_0^a(\tau,\vec{x})t^a_{R_1}}\right\rangle\nonumber\\
& & \hspace{2.5cm}\times\,\left\langle{\rm tr}\,{\cal P}\,e^{i\int_0^\beta d\tau\,A_0^a(\tau,\vec{x})t^a_{R_2}}\right\rangle\,,
\eeq
which rewrites $\ell_{R_1\otimes R_2}=\ell_{R_1}\ell_{R_2}$.

\section{Real chemical potential}\label{app:real_mu}
Although the arguments presented in this paper are general, they were tested within a particular model that evaluates the gluonic potential for the gluon field expectation value $\langle A^a_\mu\rangle$ in a particular gauge depending on a background gauge field $\bar A^a_\mu$. Both the background and the field expectation value are constant, temporal, and along the diagonal directions of the algebra:
\beq
\bar A_\mu(\tau,\vec{x}) & = & \frac{T}{g}\delta_{\mu4}\,\bar r^jt^j\,,\\
\langle A_\mu(\tau,\vec{x})\rangle & = & \frac{T}{g}\delta_{\mu4}\,r^jt^j\,.
\eeq
The glue potential $V_{\rm glue}(r^j,\bar r^j)$ depends on both $r^j$ and $\bar r^j$ but the latter should be taken equal to a specific, confining value $\bar r^j_c$ such that $V_{\rm glue}(r^j,\bar r^j_c)$ is invariant under center transformations of $r^j$.

In addition to center transformations, there are gauge transformations that allow for various possible choices of the pair $(r,\bar r_c)$. Each corresponds to choosing both $r$ and $\bar r_c$ in a so called Weyl chamber. An appropriate choice of Weyl chamber might be efficient to express additional properties of the potential. In particular, we can choose the Weyl chamber defined \cite{vanEgmond:2023lfu}
\beq
r=4\pi\sum_{k=1}^{N_c}z_k\eta_k\,,
\eeq
where the $z_k$ lie between $0$ and $1$ and sum up to $1$, and the vectors $\eta_k$ are defined in terms of the weights of the representation $\nu=1$:
\beq
\eta_1 & = & \rho_{(1)}\,,\\
\eta_2 & = & \rho_{(1)}+\rho_{(2)}\,,\\
& \dots & 
\eeq
In this Weyl chamber, charge conjugation writes
\beq
V_{\rm glue}(z_k,\bar z_k^c)=V_{\rm glue}(z_{N_c-k},\bar z_k^c)\,.
\eeq
As for the quark contribution it is such that
\beq
V_{\rm quark}(z_k,\mu)=V_{\rm quark}(z_{N_c-k},-\mu)\,.
\eeq
We have used that the confining value is invariant under charge conjugation. A related transformation is complex conjugation which writes
\beq
V_{\rm glue}(z_k,\bar z_k^c)=V^*_{\rm glue}(z^*_{N_c-k},\bar z_k^c)\,.
\eeq
As for the quark contribution it is such that
\beq
V_{\rm quark}(z_k,\mu)=V^*_{\rm quark}(z^*_{N_c-k},\mu^*)\,.
\eeq
For real chemical potential, if we want a real potential, we should choose 
\beq
z_k=z_{N_c-k}^*\,.
\eeq
This choice also ensures that the the Polyakov loops are real when the quark chemical potential is real.

\bibliographystyle{elsarticle-num}
\bibliography{refs}

@article{Aoki:2006we,
    author = "Aoki, Y. and Endrodi, G. and Fodor, Z. and Katz, S. D. and Szabo, K. K.",
    title = "{The Order of the quantum chromodynamics transition predicted by the standard model of particle physics}",
    eprint = "hep-lat/0611014",
    archivePrefix = "arXiv",
    doi = "10.1038/nature05120",
    journal = "Nature",
    volume = "443",
    pages = "675--678",
    year = "2006"
}

@article{Borsanyi:2010bp,
    author = "Borsanyi, Szabolcs and Fodor, Zoltan and Hoelbling, Christian and Katz, Sandor D and Krieg, Stefan and Ratti, Claudia and Szabo, Kalman K.",
    collaboration = "Wuppertal-Budapest",
    title = "{Is there still any $T_c$ mystery in lattice QCD? Results with physical masses in the continuum limit III}",
    eprint = "1005.3508",
    archivePrefix = "arXiv",
    primaryClass = "hep-lat",
    reportNumber = "WUB-10-11, MIT-CTP-4152",
    doi = "10.1007/JHEP09(2010)073",
    journal = "JHEP",
    volume = "09",
    pages = "073",
    year = "2010"
}

@article{Borsanyi:2013bia,
    author = "Borsanyi, Szabocls and Fodor, Zoltan and Hoelbling, Christian and Katz, Sandor D. and Krieg, Stefan and Szabo, Kalman K.",
    title = "{Full result for the QCD equation of state with 2+1 flavors}",
    eprint = "1309.5258",
    archivePrefix = "arXiv",
    primaryClass = "hep-lat",
    doi = "10.1016/j.physletb.2014.01.007",
    journal = "Phys. Lett. B",
    volume = "730",
    pages = "99--104",
    year = "2014"
}

@article{HotQCD:2014kol,
    author = "Bazavov, A. and others",
    collaboration = "HotQCD",
    title = "{Equation of state in ( 2+1 )-flavor QCD}",
    eprint = "1407.6387",
    archivePrefix = "arXiv",
    primaryClass = "hep-lat",
    reportNumber = "BNL-105928-2014-JA",
    doi = "10.1103/PhysRevD.90.094503",
    journal = "Phys. Rev. D",
    volume = "90",
    pages = "094503",
    year = "2014"
}

@unpublished{MariSurkau:2025how,
    author = "Mari Surkau, V. Tomas and Reinosa, Urko",
    title = "{Mesons, baryons and the confinement/deconfinement transition}",
    eprint = "2504.06459",
    archivePrefix = "arXiv",
    primaryClass = "hep-ph",
    month = "4",
    year = "2025"
}

@article{MariSurkau:2025xqcd,
    author = "Mari Surkau, V. Tomas and Reinosa, Urko",
    title = "{Using net quark number gain to probe the phases of QCD}",
    eprint = "2509.26178",
    archivePrefix = "arXiv",
    primaryClass = "hep-ph",
    doi = "10.1016/j.jspc.2025.100227",
    journal = "J. Subatomic Part. Cosmol.",
    volume = "4",
    pages = "100227",
    year = "2025"
}

@unpublished{MariSurkau:2025PNJLcsCF,
    author = "Mari Surkau, V. Tomas and Reinosa, Urko",
    title = "{Interplay between the chiral and deconfinement transitions from a Curci-Ferrari-improved Polyakov loop potential}",
    year = "2025",
    note = "{In preparation.}"
}

@article{Braun:2007bx,
    author = "Braun, Jens and Gies, Holger and Pawlowski, Jan M.",
    title = "{Quark Confinement from Color Confinement}",
    eprint = "0708.2413",
    archivePrefix = "arXiv",
    primaryClass = "hep-th",
    reportNumber = "HD-THEP-07-22",
    doi = "10.1016/j.physletb.2010.01.009",
    journal = "Phys. Lett. B",
    volume = "684",
    pages = "262--267",
    year = "2010"
}

@article{Dumitru:2013xna,
    author = "Dumitru, Adrian and Guo, Yun and Korthals Altes, Chris P.",
    title = "{Two-loop perturbative corrections to the thermal effective potential in gluodynamics}",
    eprint = "1305.6846",
    archivePrefix = "arXiv",
    primaryClass = "hep-ph",
    doi = "10.1103/PhysRevD.89.016009",
    journal = "Phys. Rev. D",
    volume = "89",
    number = "1",
    pages = "016009",
    year = "2014"
}

@article{Reinosa:2019xqq,
    author = "Reinosa, Urko",
    title = "{Perturbative aspects of the deconfinement transition -- Physics beyond the Faddeev-Popov model}",
    eprint = "2009.04933",
    archivePrefix = "arXiv",
    primaryClass = "hep-th",
    doi = "10.1007/978-3-031-11375-8",
    month = "6",
    year = "2022",
    journal = "Lect. Notes Phys.",
    volume = "1006",
}

@article{Aarts2023PhasePhysics,
    author = "Gert Aarts and Joerg Aichelin and Chris Allton and Andreas Athenodorou and Dimitrios Bachtis and Claudio Bonanno and Nora Brambilla and Elena Bratkovskaya and Mattia Bruno and Michele Caselle and others",
    title = "{Phase Transitions in Particle Physics}: {Results and Perspectives from Lattice Quantum Chromo-Dynamics}",
    eprint = "2301.04382",
    archivePrefix = "arXiv",
    primaryClass = "hep-lat",
    doi = "10.1016/j.ppnp.2023.104070",
    journal = "Prog. Part. Nucl. Phys.",
    volume = "133",
    pages = "104070",
    year = "2023"
}

@article{Svetitsky1982CriticalTransitions,
   author = {Benjamin Svetitsky and Laurence G. Yaffe},
   doi = {10.1016/0550-3213(82)90172-9},
   issn = {0550-3213},
   issue = {4},
   journal = {Nucl. Phys. B},
   month = {12},
   pages = {423-447},
   publisher = {North-Holland},
   title = {Critical behavior at finite-temperature confinement transitions},
   volume = {210},
   year = {1982}
}

@article{Polyakov1978ThermalLiberation,
  author =        {Polyakov, A.~M.},
  journal =       {Phys. Lett. B},
  month =         {1},
  number =        {4},
  pages =         {477--480},
  publisher =     {North-Holland},
  title =         {{Thermal properties of gauge fields and quark
                   liberation}},
  volume =        {72},
  year =          {1978},
  doi =           {10.1016/0370-2693(78)90737-2},
  issn =          {0370-2693},
}

@article{McLerran:1981pb,
    author = "McLerran, Larry D. and Svetitsky, Benjamin",
    title = "{Quark Liberation at High Temperature: A Monte Carlo Study of SU(2) Gauge Theory}",
    reportNumber = "NSF-ITP-81-08",
    doi = "10.1103/PhysRevD.24.450",
    journal = "Phys. Rev. D",
    volume = "24",
    pages = "450",
    year = "1981"
}

@article{Brown:1988qe,
    author = "Brown, F. R. and Christ, N. H. and Deng, Y. F. and Gao, M. S. and Woch, T. J.",
    title = "{Nature of the Deconfining Phase Transition in SU(3) Lattice Gauge Theory}",
    doi = "10.1103/PhysRevLett.61.2058",
    journal = "Phys. Rev. Lett.",
    volume = "61",
    pages = "2058",
    year = "1988"
}

@article{Gupta:2007ax,
    author = "Gupta, Sourendu and Huebner, Kay and Kaczmarek, Olaf",
    title = "{Renormalized Polyakov loops in many representations}",
    eprint = "0711.2251",
    archivePrefix = "arXiv",
    primaryClass = "hep-lat",
    reportNumber = "BI-TP-2007-30, BNL-NT-07-45, TIFR-TH-07-30",
    doi = "10.1103/PhysRevD.77.034503",
    journal = "Phys. Rev. D",
    volume = "77",
    pages = "034503",
    year = "2008"
}

@article{Fukushima2004ChiralLoop,
   author = {Kenji Fukushima},
   doi = {10.1016/J.PHYSLETB.2004.04.027},
   eprint = "hep-ph/0310121",
   archivePrefix = "arXiv",
   issn = {0370-2693},
   issue = {3-4},
   journal = {Phys. Lett. B},
   month = {7},
   pages = {277-284},
   publisher = {North-Holland},
   title = {Chiral effective model with the Polyakov loop},
   volume = {591},
   year = {2004}
}

@article{Reinosa:2014ooa,
    author = "Reinosa, U. and Serreau, J. and Tissier, M. and Wschebor, N.",
    title = "{Deconfinement transition in SU($N$) theories from perturbation theory}",
    eprint = "1407.6469",
    archivePrefix = "arXiv",
    primaryClass = "hep-ph",
    doi = "10.1016/j.physletb.2015.01.006",
    journal = "Phys. Lett. B",
    volume = "742",
    pages = "61--68",
    year = "2015"
}

@article{Ratti:2005jh,
    author = "Ratti, Claudia and Thaler, Michael A. and Weise, Wolfram",
    title = "{Phases of QCD: Lattice thermodynamics and a field theoretical model}",
    eprint = "hep-ph/0506234",
    archivePrefix = "arXiv",
    doi = "10.1103/PhysRevD.73.014019",
    journal = "Phys. Rev. D",
    volume = "73",
    pages = "014019",
    year = "2006"
}

@article{Roessner:2006xn,
    author = "Roessner, Simon and Ratti, Claudia and Weise, W.",
    title = "{Polyakov loop, diquarks and the two-flavour phase diagram}",
    eprint = "hep-ph/0609281",
    archivePrefix = "arXiv",
    reportNumber = "ECT*-06-16",
    doi = "10.1103/PhysRevD.75.034007",
    journal = "Phys. Rev. D",
    volume = "75",
    pages = "034007",
    year = "2007"
}

@article{Fukushima:2008wg,
    author = "Fukushima, Kenji",
    title = "{Phase diagrams in the three-flavor Nambu-Jona-Lasinio model with the Polyakov loop}",
    eprint = "0803.3318",
    archivePrefix = "arXiv",
    primaryClass = "hep-ph",
    reportNumber = "YITP-08-19",
    doi = "10.1103/PhysRevD.77.114028",
    journal = "Phys. Rev. D",
    volume = "77",
    pages = "114028",
    year = "2008",
    note = "[Erratum: Phys.Rev.D 78, 039902 (2008)]"
}

@article{Lo:2013hla,
    author = "Lo, Pok Man and Friman, Bengt and Kaczmarek, Olaf and Redlich, Krzysztof and Sasaki, Chihiro",
    title = "{Polyakov loop fluctuations in SU(3) lattice gauge theory and an effective gluon potential}",
    eprint = "1307.5958",
    archivePrefix = "arXiv",
    primaryClass = "hep-lat",
    doi = "10.1103/PhysRevD.88.074502",
    journal = "Phys. Rev. D",
    volume = "88",
    pages = "074502",
    year = "2013"
}

@article{MariavanEgmond2022ATemperature,
   author = {Duifje Maria van Egmond and Urko Reinosa and Julien Serreau and Matthieu Tissier},
   doi = {10.21468/SciPostPhys.12.3.087},
   issn = {2542-4653},
   issue = {3},
   journal = {SciPost Phys.},
   month = {3},
   pages = {087},
   title = {A novel background field approach to the confinement-deconfinement transition},
   volume = {12},
   eprint = "2104.08974",
   archivePrefix = "arXiv",
   primaryClass = "hep-ph",
   year = {2022}
}

@article{MariSurkau2024DeconfinementDependences,
    author = "Surkau, Victor Tomas Mari and Reinosa, Urko",
    title = "{Deconfinement transition within the Curci-Ferrari model: Renormalization scale and scheme dependences}",
    eprint = "2401.17869",
    archivePrefix = "arXiv",
    primaryClass = "hep-ph",
    doi = "10.1103/PhysRevD.109.094033",
    journal = "Phys. Rev. D",
    volume = "109",
    number = "9",
    pages = "094033",
    year = "2024"
}

@article{MariSurkau2025Heavy-quarkModel,
    author = "Mari Surkau, V. Tomas and Reinosa, Urko",
    title = "{Heavy-quark corner of the Columbia plot from the center-symmetric Curci-Ferrari model}",
    eprint = "2503.22373",
    archivePrefix = "arXiv",
    primaryClass = "hep-ph",
    doi = "10.1103/7dx2-vmj6",
    journal = "Phys. Rev. D",
    volume = "112",
    number = "1",
    pages = "014002",
    year = "2025"
}

@article{Fromm2012ThePotentials,
    author = {Michael Fromm and Jens Langelage and Stefano Lottini and Owe Philipsen},
    title = {{The QCD deconfinement transition for heavy quarks and all baryon chemical potentials}},
    journal = {J. High Energy Phys.},
    year = {2012},
    month = {Jan},
    day = {11},
    volume ={01},
    pages = {42},
    doi = {10.1007/JHEP01(2012)042},
    issn = {1029-8479},
    eprint = "1111.4953",
    archivePrefix = "arXiv",
    primaryClass = "hep-lat"
}

@misc{Gernot,
    author = "Eichmann, Gernot",
    title = "{QCD and Hadron Physics}",
    journal = "Lecture notes",
    year = "2020",
    howpublished = {\url{http://cftp.ist.utl.pt/~gernot.eichmann/2020-QCDHP/App-SU(N).pdf}},
    note = {{Lecture notes, Accessed: 26.11.2025}}
}

@article{Lucini:2012gg,
    author = "Lucini, Biagio and Panero, Marco",
    title = "{SU(N) gauge theories at large N}",
    eprint = "1210.4997",
    archivePrefix = "arXiv",
    primaryClass = "hep-th",
    reportNumber = "HIP-2012-24-TH, NSF-KITP-12-190",
    doi = "10.1016/j.physrep.2013.01.001",
    journal = "Phys. Rept.",
    volume = "526",
    pages = "93--163",
    year = "2013"
}

@article{vanEgmond:2023lfu,
    author = "van Egmond, Duifje Maria and Reinosa, Urko",
    title = "{Center-symmetric Landau gauge: Further signatures of confinement}",
    eprint = "2309.12699",
    archivePrefix = "arXiv",
    primaryClass = "hep-th",
    doi = "10.1103/PhysRevD.109.036002",
    journal = "Phys. Rev. D",
    volume = "109",
    number = "3",
    pages = "036002",
    year = "2024"
}

\end{document}